\documentclass[9pt,twocolumn,twoside]{osajnl}

\journal{josaa} 

\setboolean{shortarticle}{false} 

\title{Wavefront retrieval through random  pupil-plane phase probes: Gerchberg-Saxton approach}

\author[1,2,*]{Eugene Pluzhnik}
\author[1,2]{Dan Sirbu}
\author[1]{Ruslan Belikov}
\author[1]{Eduardo Bendek}
\author[3,\textdagger]{Vladimir N. Dudinov}

\affil[1]{NASA Ames Research Center, Moffett Field, CA 94035}
\affil[2]{Universities Space Research Association, 7178 Columbia Gateway Dr., Columbia, MD 21046}
\affil[3]{Institute of Astronomy, Kharkiv National University, 35 Sumska Str., Kharkiv, 61022, Ukraine\newline \textdagger Deceased 2 May 2016}
\affil[*]{Corresponding author: yevgeniy.a.pluzhnyk@nasa.gov}

\ociscodes{(100.5070) Phase retrieval; (110.1080) Active or adaptive optics;(010.7350) Wave-front sensing.}

\begin{abstract}
A pupil-plane wavefront reconstruction  procedure is proposed based on analysis of a sequence of 
focal-plane images  corresponding random pupil-plane phase probes. The developed method 
provides the unique nontrivial solution of wavefront retrieval problem and
shows
global convergence to this solution demonstrated using a Gerchberg-Saxton
implementation. The  method is general and
can be used in any optical system 
equipped with a wavefront control device.
The uniqueness of a global solution for such pupil-plane perturbations is proven and conditions for algorithm convergence to the unique solution are discussed. The presented numerical simulation and lab experimental results show
low noise sensitivity, high  reliability and robustness  of the proposed approach
for high quality optical wavefront restoration.
Laboratory experiments have shown $\lambda$/14 rms accuracy in retrieval of a
poked deformable mirror  actuator
fiducial pattern with spatial resolution of 20-30$~\mu$m
that is comparable with accuracy of direct high-resolution interferometric
measurements.

\end{abstract}

\setboolean{displaycopyright}{true}
\begin{document}

\maketitle

\section{Introduction}
Optical applications often require knowledge of both phase and amplitude
of an optical wavefront. Similar problems 
arise in various fields such as electron microscopy \cite{Saxton_1978},
X-ray crystallography  \cite{Millane_1990}, astronomy \cite{Dainty_1987},  optical imaging \cite{Shechtman_2015}, etc. 
The particular interest 
of the authors of this paper is
inspired by the importance  
of optical phase measurements for
direct exoplanet imaging, where wavefront reconstruction is needed both for accurate modeling of the instrument, as well as for wavefront control methods such as Electric Field Conjugation (EFC) \cite{Giveon_2007, Giveon_2007_1}.


Since it is not always possible to measure the desired wavefront directly,
different methods have been developed that allow its reconstruction based on intensity measurements only. Phase information is absent from intensity-only measurements \cite{ONeil_1963} and cannot be reconstructed unless some technique is used that encodes phase information in a sequence of intensity images in some controlled fashion.

It was found in earlier works \cite{Bruck_1979, Bates_1982, Hayes_1982, Fienup_1984} that for almost every two-dimensional pupil-plane wavefront, a unique solution exists if the focal-plane intensities are known together with some constraints applied to the pupil aperture. 
In this case the wavefront reconstruction problem can be formulated as the problem of a complex-valued signal reconstruction from the modulus of its Fourier transform. However, even if a unique solution exists, it is not always possible to find this solution, due to convergence issues associated with wavefront reconstruction algorithms \cite{Levi_1984, Fienup_1986, Chretien_1996}. 
The solution uniqueness is not absolute, because a few trivial ambiguities still remain unsolved and can affect algorithm convergence \cite{Fannjiang_2012} not counting the 
non-convexity of the Fourier magnitude constraint \cite{Levi_1984, Bauschke_2002}.

\begin{figure}[t]
\begin{center}
\begin{tabular}{c}
\includegraphics[width=1.02\hsize]{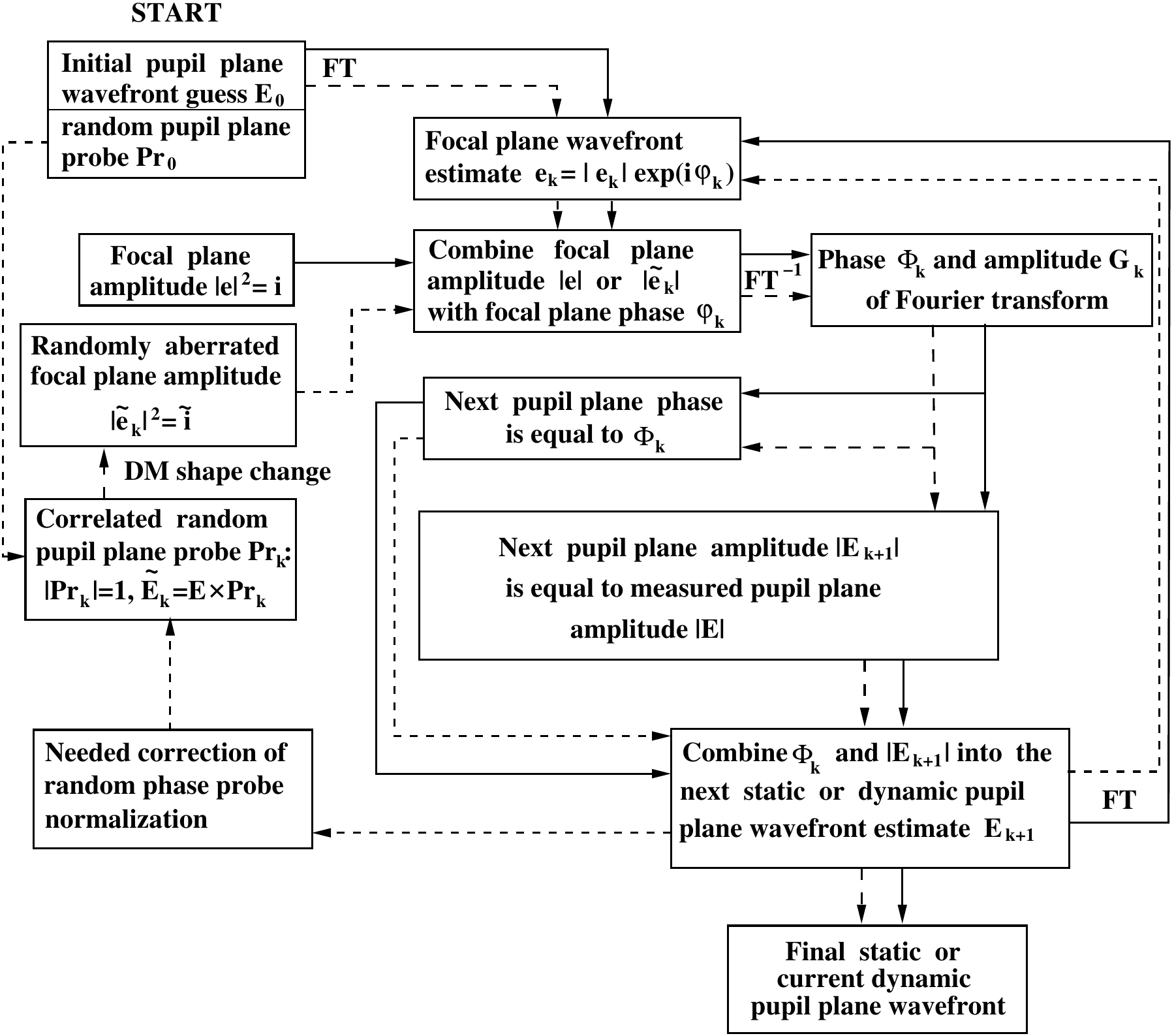}
\end{tabular}
\end{center}
\caption[fig1]
{\label{fig1}
Schematic of phase retrieval algorithms: Gerchberg-Saxton iterative loop
is shown with solid lines, the random ~ phase ~ probes ~ modification is ~
indicated with  dashed~ lines. Direct and ~ inverse Fourier ~ transforms are
marked~ as~ ``FT'' and~ ``FT$^{-1}$''.
~ Correlated~ random~ pupil-plane ~ probes~
can be~ produced~ by changing~ the DM shape.}
\end{figure}

In this paper we propose a pupil-plane wavefront reconstruction method based on analysis of a sequence of randomly aberrated focal-plane images (Section \ref{method})  produced by the Deformable Mirror (DM)  or other means.
Random phase aberrations 
in the pupil plane provide a unique (Section~\ref{theorem}) and globally converging solution (Section~\ref{methodc}) for measuring pupil-plane wavefront as demonstrated through simulations and experimental results described in Sections \ref{simulations} and \ref{experiment}. The low noise sensitivity, fast convergence of the applied algorithms to the global solution in the presence of both static and dynamic pupil-plane aberrations (as described in Section \ref{simulations}), high reliability and robustness,
insensitivity to phase discontinuities and non-common path errors, make the proposed approach useful in a wide range of optical applications where high quality wavefront retrieval is needed (as discussed in Section \ref{discussion}).

\section{Method}
\label{method}

Historically, the first successful phase retrieval method was proposed by Gerchberg and Saxton \cite{Gerchberg_1972}. The algorithm was originally developed to restore the pupil-plane wavefront assuming that both pupil-plane intensities $I({\bf u})=|E({\bf u)}|^2$ and
focal-plane intensities
$i({\bf r})=|e({\bf r})|^2$ are known.
The pupil-plane (complex-valued) electric field  E({\bf u}) and the focal-plane electric field  $e({\bf r})$
are related by the Fourier transform
\begin{eqnarray}
e({\bf r})=~|e({\bf r})|\exp[i\varphi({\bf r})]~~~=\int E({\bf u}) \exp[i2\pi{\bf u}\cdot{\bf r}] d{\bf u}, \nonumber \\
E({\bf u})=|E({\bf u})|\exp[i\Phi({\bf u})]=\int e({\bf r}) \exp[-i2\pi{\bf u}\cdot{\bf r}] d{\bf r},
\label{eq1}
\end{eqnarray}
where $|E({\bf u)}|$ and $|e({\bf r})|$ are wavefront amplitudes,
$\Phi({\bf u})$ and $\varphi({\bf r})$ are wavefront phases, and
${\bf u}$ and ${\bf r}$ are radius-vectors in the pupil plane and the focal plane
respectively. A constant phase factor as well as a normalizing factor $1/\lambda f$, where
$\lambda$ is the wavelength and $f$ is the system  focal length,
are not included in Eq.~\ref{eq1} for simplicity.

The  Gerchberg-Saxton iterative loop switches between pupil and focal plane performing the following sequence of steps
in the $(k+1)$-th iteration  (the schematic of all described algorithms is shown in Fig.~\ref{fig1}):
\begin{enumerate}
\item Fourier transform the current wavefront estimate $E_k({\bf u})$
      \begin{equation} e_k({\bf r})=|e_k({\bf r})|\exp[i\varphi_k({\bf r})]=
      \int E_k({\bf u}) \exp[i2\pi{\bf u}\cdot{\bf r}] d{\bf u}.
      \end{equation}
\item Replace the amplitude $|e_k({\bf r})|$ of the resulting Fourier transform with the
      measured focal-plane amplitude $|e({\bf r})|$
      \begin{equation}
	    g_k({\bf r})=|e({\bf r})|\exp[i\varphi_k({\bf r})].
      \end{equation}
\item Inverse Fourier transform the current focal-plane wavefront estimation  $g_k({\bf r})$
      \begin{equation} G_k({\bf u})=|G_k({\bf u})|\exp[i\Phi_k({\bf u})]=
      \int g_k({\bf r}) \exp[-i2\pi{\bf u}\cdot{\bf r}] d{\bf u}.
      \end{equation}
\item Obtain the next pupil-plane wavefront estimate by replacing  the amplitude $|G_k({\bf u})|$
      with the measured pupil-plane amplitude $|E({\bf u})|$
      \begin{equation}
	    E_{k+1}({\bf u})=|E({\bf u})|\exp[i\Phi_k({\bf u})].
      \end{equation}
\end{enumerate}

Although pupil-plane and focal-plane intensity distributions 
used in an iterative loop as constraints can provide 
efficient locally converging algorithms
\cite{Gerchberg_1972, Fienup_1982, Levi_1984}. they cannot guarantee 
global convergence to a unique solution even when such a solution exists
\cite{Seldin_1990, Levi_1984}.
As a result projection algorithms, which include the Gerchberg-Saxton method, stagnate near the closest local minimum instead
(see examples in 
\cite{Fiddy_1983,Fannjiang_2012a}
).
The main reasons for the stagnation are the existence of ambiguous (non-unique)
pupil-plane wavefronts \cite{Seldin_1990} and
the non-convexity of the Fourier magnitude  constrain
\cite{Levi_1984, Bauschke_2002}. Even trivial ambiguities such as global tip/tilt or conjugate inversion of the wavefront 
in combination with loose or symmetric wavefront support can cause algorithm stagnation
\cite{Fienup_1986, Fienup_1987, Fannjiang_2012, Fannjiang_2012a} 
near a solution where
the shifted wavefront appears combined with the recovered wavefront. Modifications of the Gerchberg-Saxton algorithm such as basic input-output and hybrid
input-output algorithms by Fienup \cite{Fienup_1978, Fienup_1982} improve
convergence to the global solution in some cases but still demonstrate stagnation
in other cases \cite{Fienup_1986, Seldin_1990, Fannjiang_2012a}.

Another group of wavefront retrieval algorithms are ``diversity'' algorithms,
which analyze focal-plane (or some intermediate plane) 
intensity variations caused by predetermined changes of the pupil plane to measure the pupil-plane wavefront.
The desired ``diversity'' can be produced by either pupil-plane phase
\cite{Schiske_1968, Misell_1973, Misell_1973a, Gonsalves_1982}
or amplitude \cite{Walker_1981, Gonsalves_1986, Dudinov_1988} changes 
that provide
the unique solution for the wavefront
retrieval. It has been shown \cite{Pluzhnik_1996}
that analytically unique (up to the global phase)
2-dimensional pupil-plane wavefront solution can be found by using three different exponential
pupil-plane amplitude probes.
Similar conclusions  have also been made for 
phase retrieval by using phase probes  \cite{Borde_2006,Fannjiang_2012}. 
Pupil-plane wavefront diversity removes the main reason for wavefront retrieval stagnation, namely non-uniqueness of the solution.
Though non-convexity of the constraints can still be theoretically relevant to some open-loop phase wavefront diversity applications, the related numerical problems can be easily solved by changing the predetermined diversities.
As a result, diversity algorithms are faster and show better global convergence
compared with Gerchberg-Saxton based algorithms.
The simplest phase diversity may be produced by defocusing (similar to the classical Hartmann optical test) that introduces a quadratic phase
variation across the pupil. Such diversity, however, is not very sensitive
to non-symmetric phase aberrations \cite{Smith_2004}.
More complicated diversity methods often need additional optics which may cause non-common path errors.
Phase diversities need to be well calibrated \cite{Paul_2013} and often require perfect knowledge of optical
system parameters \cite{Lamb_2015}.

Phase diversity can be combined with projective algorithms to improve  
the quality of wavefront reconstruction. Although this approach is often useful
in stable laboratory conditions \cite{Guyon_2010_1}, it cannot be easily adapted to optical aberrations 
randomly changing with time. 
A classical example of such aberrations is introduced by the Earth-turbulent atmosphere, which serves as an inspiration for our approach.

We consider the problem of restoration of the pupil-plane phase caused by the atmosphere turbulence in more detail.
We assume that the pupil-plane wavefront
$E({\bf u},t)=|E({\bf u},t)|\exp[i\Phi({\bf u},t)]$
is temporally changing under the influence of the atmosphere turbulence and we try to estimate $E({\bf u},t_k)$, where $\{t_k\}_{k=1}^\infty$ is a uniform time sequence, by using the measured sequence of focal-plane intensities $i({\bf r},t_{k})=|e({\bf r},t_{k})|^2$. We also assume the pupil-plane amplitude  $|E({\bf u},t)|=|E({\bf u})|$ is not changing with time being constant across the telescope pupil and equal to 0 outside.
Consider the  following modification of the Gerchberg-Saxton algorithm: 
 

\begin{enumerate}
\item Fourier transform the current wavefront estimate
      obtained in the time moment $t_k$
      \begin{eqnarray} e_k({\bf r},t_k)=|e_k({\bf r},t)|\exp[i\varphi_k({\bf r},t_k)]=\nonumber \\
      =\int E_k({\bf u},t_k) \exp[i2\pi{\bf u}\cdot{\bf r}] d{\bf u}.
      \end{eqnarray}
\item Replace the amplitude $|e_k({\bf r},t_k)|$ of the resulting Fourier transform
      with the
      new focal-plane amplitude $|e({\bf r},t_{k+1})|$ that is changing with time
      \begin{equation}
	    g_k({\bf r},t_{k+1})=|e({\bf r},t_{k+1})|\exp[i\varphi_k({\bf r},t_k)].
      \end{equation}
\item Inverse Fourier transform the current focal-plane wavefront estimation  $g_k({\bf r},t_{k+1})$
      \begin{eqnarray} G_k({\bf u}, t_{k+1})=|G_k({\bf u},t_{k+1})|\exp[i\Phi_k({\bf u},t_{k+1})]=\nonumber \\
      =\int g_k({\bf r},t_{k+1}) \exp[-i2\pi{\bf u}\cdot{\bf r}] d{\bf u}.~~~~~~~
      \end{eqnarray}
\item Obtain the next pupil-plane wavefront estimate by replacing the amplitude $|G_k({\bf u}),t_{k+1}|$
      with the telescope pupil-plane amplitude $|E({\bf u})|$
      \begin{equation}
	    E_{k+1}({\bf u},t_{k+1})=|E({\bf u})|\exp[i\Phi_k({\bf u},t_{k+1})].
      \end{equation}
\end{enumerate}

It is clear that in the case of completely uncorrelated atmospheric wavefronts algorithm never converges. However,
if the successive wavefront change is small enough to provide 
continuous wavefront transformation, one can expect the sequence
$E_{k}({\bf u},t_{k})$ to converge to some temporally changing  solution. This solution is expected to be negligibly different from the
actual pupil-plane wavefront $E({\bf u},t_{k})$ . The accuracy of such restoration should be limited by the rms of the successive wavefronts phase difference.

\section{Uniqueness Theorem}
\label{theorem}
The described modification of the Gerchberg-Saxton algorithm 
can only converge to a unique, non-trivial solution. 
This solution is a continuous time sequence of reconstructed pupil-plane phase distributions, each of which coincides with the corresponding atmospheric wavefront.
By non-trivial solutions we mean all solutions
except for those whose difference is caused by the global wavefront tip/tilt and piston. 
The phase solution ambiguity  caused by the ``conjugation'' symmetry of the wavefront $E({\bf u})$ and its complex conjugate $E^\ast({-\bf u})$ is taken into account in our future discussion. 
The solution uniqueness directly follows from the following lemmas. 


\begin{description}
\item[\bf Lemma 1:] Let $E_1({\bf u})$ and $E_2({\bf u})$ be two different pupil-plane wavefronts 
that form the identical intensity distribution in the focal plane. Any small enough pupil-plane phase 
perturbation $\Delta\Phi({\bf u})$ applied to  $E_1({\bf u})$ and $E_2({\bf u})$ transforms corresponding focal-plane
intensities $I_1({\bf r})$ and $I_2({\bf r})$
such that they no longer coincide with each other except for some special cases of cross-symmetry between the wavefronts. 
\end{description}
{\bf Proof:} The m-th intensity distribution $I_m({\bf r})$
is related to the focal-plane electric field
$e_m({\bf r})$ as $I_m({\bf r})=|e_m({\bf r})|$.
According to the lemma assumption, $I_1({\bf r})=I_2({\bf r})$. The perturbation $\Delta\Phi({\bf u})$ applied to the m-th ($m=1,2$) pupil-plane wavefront transforms
$E_m({\bf u})$ into  $E_m({\bf u})\exp[i\Delta\Phi({\bf u})]$,
$\,\,e_m({\bf r})$ into $e_{\Delta\Phi,m}\,({\bf r})$, and   $I_m({\bf r})$ into $I_{\Delta\Phi,m}\,({\bf r})$.
In discrete form, the focal-plane intensity change caused by this perturbation can be expressed as a linear combination of  $\Delta\Phi({\bf u}_k)$
values taken at the $k$-th sampling point in the pupil plane
\begin{equation}
\Delta I_{\Delta\Phi,m}({\bf r}_j) = I_{\Delta\Phi,m}({\bf r}_j)-I_l({\bf r}_j)=\sum_k B^{(m)}_{jk}\Delta\Phi({\bf u}_k), 
\end{equation}
where ${\bf r_j}$ is the $j$-th sampling point in the focal plane,
elements of   matrix $B^{(m)}=[\,B^{(m)}_{jk}\,]$ are equal to $\partial I_m({\bf r}_j)/{\partial \Phi_m({\bf u}_k)}$, $\Phi_m({\bf u}_k)=\arg [E_m({\bf u}_k)]$, and the summation is performed for each sampling point in the pupil.


Let $K$ be the total number  of sampling points over the pupil aperture.
To make intensities $I_{\Delta\Phi,1}\,({\bf r})$
and $I_{\Delta\Phi,2}\,({\bf r})$ equal to each other
the perturbation $\Delta\Phi({\bf u})$ should be a solution of the following system of linear equations
\begin{equation}
      \sum_{k=1}^K (B^{(1)}_{jk}-B^{(2)}_{jk})\, \Delta\Phi({\bf u}_k)=0, \mbox{    } j=1, 2, ... K,
\label{uniq2}
\end{equation}
written for $K$ independent sampling points in the image plane. Let $[L_{jk}]=[B^{(1)}_{jk}-B^{(2)}_{jk}]$ be the matrix of the linear system~(\ref{uniq2}). Two different cases are possible.

In case $\det\, [L_{jk}]\neq 0$, the system~(\ref{uniq2}) has a unique solution $\Delta\Phi({\bf u})=0$ ~ that makes equality $I_{\Delta\Phi,1}\,({\bf r})=I_{\Delta\Phi,2}\,({\bf r})$ impossible in some continuous 
area of small arbitrary changes of the perturbation 
$\Delta\Phi({\bf u})$ except for $\Delta\Phi({\bf u})=0$.

The case $\det\, [L_{jk}] = 0$ corresponds to 
infinitely many solutions of the system~(\ref{uniq2}). 
All these solutions belong to some continuous area of changes of perturbation $\Delta\Phi({\bf u})$, but they are not arbitrary. 
``Cross-symmetry'' between $E_1({\bf u})$ and $E_2({\bf u})$ occurs when $\det\ [L_{jk}] = 0$ and is thus associated with an infinite set of solutions. Thus, Lemma 1 is considered proven. All possible cases
of cross-symmetry between $E_1({\bf u})$ and $E_2({\bf u})$ are analyzed later. $\Box$ 

\begin{description}
\item [\bf Corollary 1: ] One direct consequence immediately follows from Lemma~1.
Let two different pupil-plane wavefronts $E_1({\bf u})$ and $E_2({\bf u})$ be influenced by a continuous phase perturbation $\Delta\Phi_\gamma({\bf u})$ 
such that $\det [L_{jk}^\gamma]\neq 0$ 
at any point of the transformation. The perturbation $\Delta\Phi_\gamma({\bf u})$ is given here in the form of a continuous ``transformation line'' 
parametrized by  $\gamma\in [ \, 0,1 ] \,$ and the matrix $[L_{jk}]$ at the point $\gamma$ is denoted as  $[L_{jk}^\gamma]$. 
In this case,  pupil-plane wavefronts $E_1({\bf u})\exp[i\Delta\Phi_\gamma({\bf u})]$ and $E_2({\bf u})\exp[i\Delta\Phi_\gamma({\bf u})]$ produce  different focal-plane images $I_{\gamma,1}({\bf r})\neq I_{\gamma,2}({\bf r})$ at any point $\gamma$ of the transformation line except for a possible finite set of points $\{\gamma_n\}$
where $I_{\gamma_n\, ,1}({\bf r})=I_{\gamma_n\, ,2}({\bf r})$.
\end{description} 
{\bf Proof:} The assumption that 
the set $\{\gamma_n\}$ is infinite means that there exists a converging
subset $\{\gamma_{n_l}\}$ such that 
linear system~(\ref{uniq2}) has only one solution for each element of the subset $\{\gamma_{n_l}\}$. 
Let $\gamma_c$ be the point of convergence of $\{\gamma_{n_l}\}$. This implies that 
$\gamma_c\in [0,1]$ and 
in an arbitrary small neighborhood around $\gamma_c$ indefinitely many solutions of the linear system~(\ref{uniq2}) exist. Therefore, $\det [L_{jk}^\gamma]=0$ contradicts the assumption $\det [L_{jk}^\gamma]\neq 0$. The obtained contradiction proves the corollary.$\Box$ 

The linear system~(\ref{uniq2}) always has at least one solution. The analysis for the case of infinitely many solutions is given below. 
It is generally accepted that all non-trivial ambiguous phase solutions fall under one of the following three cases of pupil-plane wavefront symmetry:


\noindent {\bf Case 1}. The case of the ``convolution'' symmetry for which the focal-plane wavefront $e({\bf r})$ can be presented as a factor of two entire functions $e({\bf r})=f({\bf r})g({\bf r})$ \cite{Hayes_1982}.  The corresponding pupil-plane wavefront $E({\bf u})$ is a convolution of two  functions $E({\bf u})=F({\bf u}) \otimes G({\bf u})$, where $F({\bf u})$ and  $G({\bf u})$
are Fourier transforms of functions
$f({\bf r})$  and $g({\bf r})$, and 
symbol ``$\otimes$'' is used to denote the convolution operation. 
In this case the pupil-plane wavefronts $F({\bf u}) \otimes G({\bf u}), \, F({\bf u}) \otimes G^\ast ({-\bf u}),  F^\ast (-{\bf u}) \otimes G({\bf u}) $, and  $F^\ast (-{\bf u}) \otimes G^\ast (-{\bf u}) $ form identical  focal-plane images  coinciding with the focal-plane intensity $I({\bf r})$. 


\noindent {\bf Case 2}. In the case of the conjugation symmetry the pupil-plane wavefront  $E_2({\bf u})=E_1^\ast (-{\bf u})$ always produces the focal-plane intensity $I_2({\bf r})$ that is equal to
the focal-plane intensity $I_1({\bf r})$ produced by the wavefront
$E_1({\bf u})$.

\noindent {\bf Case 3}. The case of separable variables for which the pupil-plane wavefront $E({\bf u})$ can be presented as a product of two factors $E({\bf u})=F({u_x}) \, G(u_y)$, 
$F({u_x})$ and $G({u_y})$ are functions of one variable, and  $u_x$ and $u_y$ are components of the vector ${\bf u}$.
In this case the pupil-plane wavefronts $F({u_x}) \, G(u_y)$,  $F({u_x}) \, G^\ast (-u_y),  F^\ast ({-u_x}) \, G(u_y) $, and  $F^\ast ({-u_x}) \, G^\ast (-u_y) $ also produce identical focal-plane intensities coinciding with the
 focal-plane intensity $I({\bf r})$.

We call all solutions  described in Cases 1--3 ``conjugated''
wavefront solutions because  they are all related to the complex conjugation of one of the function which describes the wavefront.


It is clear that any continuous pupil-plane wavefront transformation, which preserves its symmetry, that may keep multiple solutions.
We consider such transformations in detail next. For simplicity we do not analyze cases 
of mixed symmetry. 


\begin{description}
\item [\bf Lemma 2:] Let a pupil-plane  wavefront $E_0({\bf u})$ be
a convolution of functions $F_0({\bf u})$ and  $G_0({\bf u})$,
and a pupil-plane  wavefront $E_1({\bf u})$ be
a convolution of functions $F_1({\bf u})$ and  $G_1({\bf u})$.
Consider all possible continuous phase transformations of the pupil \, $\exp[i \Delta\Phi_{\gamma} ({\bf u})]$
that convert
 $E_0({\bf u})=F_0({\bf u}) \otimes  G_0({\bf u})$
into $E_1({\bf u})=F_1({\bf u}) \otimes  G_1({\bf u})$
such that $E_\gamma ({\bf u})$ can be presented as a convolution 
at any point $\gamma$ of the transformation line, i.e.
\begin{equation}
E_\gamma ({\bf u})=E_0 ({\bf u}) \, \exp[i \Delta\Phi_{\gamma} ({\bf u})]=F_\gamma ({\bf u}) \otimes  G_\gamma ({\bf u}),
\label{uniq3}
\end{equation}
where $F_\gamma ({\bf u})$ and $G_\gamma ({\bf u})$ are some functions. Lemma 2 asserts that such transformations does not exist.
\end{description}
{\bf Proof:} To preserve the continuity of $E_\gamma ({\bf u})$ along the transformation line  
both functions $F_\gamma({\bf u})$ and  $G_\gamma ({\bf u})$ must be continuous.
Arbitrary continuous transformations connecting $F_0({\bf u})$
with $F_1({\bf u})$, and $G_0({\bf u})$ with $G_1({\bf u})$
can be expressed in a polynomial form as  
\begin{align}
F_\gamma ({\bf u})=(1-\gamma)^n F_0({\bf u})+\sum_{k=1}^{n-1}\gamma^k (1-\gamma)^{n-k} f_k({\bf u}) + \gamma^n F_1({\bf u}), \mbox{~~} \nonumber \\
G_\gamma ({\bf u})=(1-\gamma)^n G_0({\bf u})+\sum_{k=1}^{n-1}\gamma^k (1-\gamma)^{n-k} g_k({\bf u}) + \gamma^n G_1({\bf u}), \nonumber \\
\raisetag{2\normalbaselineskip}
\label{approx}
\end{align}
where $\{f_n({\bf u})\}$ and $\{g_n({\bf u})\}$ are arbitrary
sets of functions, and $n$ is the degree of  approximating polynomials.  To simplify expressions we confine ourselves to the case 
\begin{eqnarray}
F_\gamma ({\bf u})=(1-\gamma) F_0({\bf u})+\gamma F_1({\bf u}),~  \nonumber\\
G_\gamma ({\bf u})=(1-\gamma) G_0({\bf u})+\gamma G_1({\bf u}).
\end{eqnarray}
We also suggest that $|E_\gamma ({\bf u})|^2=1$ for any value of $\gamma$.
With the simplifications adopted the equation $|E_\gamma ({\bf u})|^2=1$ can be written down in the form
\begin{eqnarray}
|(1-\gamma)^2 \, F_0 ({\bf u}) \otimes G_0({\bf u})+\gamma(1-\gamma)\times  [F_1 ({\bf u}) \otimes G_0({\bf u})\nonumber + \\+F_0 ({\bf u}) \otimes G_1({\bf u})] +\gamma^2 \, F_1 ({\bf u}) \otimes G_1({\bf u}) |^2=1.~~~~~~~~~~~~~~~~~~~~~~ 
\label{uniq4}
\end{eqnarray}
Eq.~\ref{uniq4} reflects the independence of the  wavefront amplitude
$|E_\gamma ({\bf u})|$ from the pupil-plane phase perturbation.
This is an algebraic equation (with respect to $\gamma$) of 4-th order with real coefficients. The equation has 2 or 4 real roots only two of which can belong to the interval $(0,1)$.
This means that there are no more than two points $\gamma\in (0,1)$ in which $E_\gamma ({\bf u})$ can be presented as a convolution of two functions $F_\gamma ({\bf u}) \otimes  G_\gamma ({\bf u})$.
An increase in the degree $n$ increases the degree of the polynomial~(\ref{uniq4}).
This in turn increases the number of polynomial roots that can be responsible for the convolution caused solution ambiguity 
along the transformation line. The set $\{\gamma_k\}_{k=1}^{4n-2}$ of such roots,
however, is always countable and cannot cover a continual set $[0,1]$.
In a limit where $n\rightarrow\infty$ polynomial~(\ref{uniq4}) converts into a function that allows analytical extension into the  whole complex plane $\bf C$. This is a holomorphic function that can have an infinite number of countable zeros $\{\gamma_k\}_{k=1}^\infty$ on a real axis. However, only a finite number of these zeros can be located on the segment $[0,1]$ as it follows from the uniqueness of the analytical extension theorem. As a result any continuous pupil-plane phase transformation $E_\gamma ({\bf u})$ contains only a finite number of isolated points $\{\gamma_k\}$ where  $E_{\gamma_k} ({\bf u})=F_{\gamma_k} ({\bf u}) \otimes  G_{\gamma_k} ({\bf u})$. The changes that must be included in the proof in the case of transformation line derivative discontinuities are obvious. $\Box$ 

Lemmas 1 and 2 make it possible to formulate and prove a uniqueness theorem. 

\begin{description}
\item{\bf Uniqueness theorem:} Let $E_\gamma ({\bf u})$ be a continuous sequence of pupil-plane wavefronts (parametrized by $\gamma$, $\gamma\in [0,1]$) that produces a continuous sequence of focal-plane intensities $I_\gamma ({\bf u})$. Let $|E_\gamma ({\bf u})|$ be constant with the change of $\gamma$. 
Assuming no piston term, let the focal-plane wavefront $e_\gamma({\bf r})$ not be a real function. In this case the problem of restoration of the $E_\gamma ({\bf u})$ sequence from the $I_\gamma ({\bf u})$ sequence has:

(1) infinite number of different non-trivial (no tip/tilt and piston) solutions in the case where $E_\gamma ({\bf u})$ can be presented as a product of three factors $E_\gamma ({\bf u})=G(u_y) \, F({u_x}) \exp[i \Delta\Phi_{\gamma} ({\bf u}_x)]$ or $E_\gamma ({\bf u})=F({u_x}) \, G(u_y)\exp[i \Delta\Phi_{\gamma} ({\bf u}_y)]$;

(2) four different non-trivial solutions in the case where $E_\gamma ({\bf u})$ can be presented as a product of two factors $E_{\gamma} ({\bf u})=F_{\gamma}({u_x}) \, G_{\gamma}(u_y)$
and does not allow factorization described in Case 1.

(3) two different solutions in all other cases.
\end{description}
{\bf Proof:} All non-trivial multiple wavefront  solutions  can only be  caused by conjugation related wavefront symmetries  described earlier as Cases 1--3. 
It is easy to prove analytically that the same phase perturbation $\Delta\Phi_\gamma({\bf u})$ applied to the conjugated wavefronts $E_1({\bf u})=E_\gamma ({\bf u})$ and $E_2({\bf u})=E_\gamma^\ast ({-\bf u})$ 
transforms the corresponding focal-plane images such
that $I_{\Delta\Phi_\gamma,1}({\bf r})\neq I_{\Delta\Phi_\gamma,2}({\bf r})$, which is equivalent to the fulfillment of conditions
$\det [L^\gamma_{jk}]\neq 0$ for any point $\gamma\in [0,1]$ for the matrix $[L^\gamma_{jk}]=[B^{(E_\gamma)}_{jk}-B^{(E_\gamma^\ast)}_{jk}]$ (Lemma 1). 
The condition $\det [L^\gamma_{jk}]\neq 0$ means that the wavefront conjugation  at any point $\gamma$  should be consistent with the conjugation of all other points of the transformation (Corollary 1). The actual solution at any transformation point  can be distinguished between  conjugated local (related to any single point $\gamma$) solutions if the conjugation of the actual
solution is known in at least one point of the transformation. 
If the conjugation of the actual solution is unknown, the conjugation "sign" can be 
arbitrarily chosen at some  point of the transformation  that sets conjugation of solutions at all other points. As a result 
the total number of possible wavefront solutions that provide
the observed continuous transformation of the focal-plane intensity is equal to the minimal number of  conjugated 
local wavefront solutions.  

Lemma 1 can be applied to all points where $\det [L^\gamma_{jk}]\neq 0$. At points where $\det [L^\gamma_{jk}]= 0$, a unique continuous extension of the solution becomes impossible.  In the case where variables are not separable,  $\det [L^\gamma_{jk}]$ is equal to 0 only if $E_{\gamma} ({\bf u})=E_{\gamma}^\ast ({-\bf u})$ which is equivalent to the constraint $e_\gamma ({\bf r})$ not be a real function.

We now consider each of the three theorem cases.


1. In the case where at any point
$E_\gamma ({\bf u})=G(u_y) \, F({u_x}) \exp[i \Delta\Phi_{\gamma} ({\bf u}_x)]$ the problem can be reduced to two one-dimensional problems, namely, the restoration of the function $F({u_x}) \exp[i \Delta\Phi_{\gamma} ({\bf u}_x)]$ from the modulus of the corresponding Fourier transform  $|f_\gamma({x})|$,
and the restoration of the function $G({u_y})$ from the modulus of  the corresponding  Fourier transform  $|g_({y})|$. Note that in this case $F({u_x}) \exp[i \Delta\Phi_{\gamma} ({\bf u}_x)]$ and
$|f_\gamma({x})|$ are continuously changing while
$G(u_y)$ and $|g_({y})|$ do not change during the wavefront transformation.
The follows from the third case  of the theorem the first problem has exactly two solutions.
The second problem has infinite number of solutions \cite{Bruck_1979, Dainty_1979} which determines the total number of solutions in the first case of the theorem.

2. In contrast to the first case,  in the second case  $E_{\gamma} ({\bf u})=F_{\gamma}({u_x}) \, G_{\gamma}(u_y)$, and both $F_{\gamma}({u_x})$ and  $G_{\gamma}({u_x})$ are continuously changing along the transformation line. In accordance with the Case 3 description it means that 
this case has exactly 4 solutions. They are 
$F_{\gamma}({u_x}) \, G_{\gamma}(u_y), \, F_{\gamma}({u_x}) \, G_{\gamma}^\ast (-u_y),  F_{\gamma}^\ast ({-u_x}) \, G_{\gamma}(u_y) $, and  $F_{\gamma}^\ast ({-u_x}) \, G_{\gamma}^\ast (-u_y)$. As it was discussed above  
if the actual conjugation is  known at an arbitrary point $\gamma_0$ of the phase transformation this conjugation 
can be expanded to all other points of the phase transformation.


3. There are two different reasons that are responsible for
multiple solutions in the third case. The first one is the ``convolution'' symmetry discussed in detail in the Case 3 description  and the Lemma 2.
To retain such multiple solutions along the transformation line
the ``convolution'' symmetry should be preserved at every point of the transformation. However, this is impossible in according to Lemma 2. Thus, only the conjugation symmetry of  the wavefront causes multiple solutions in this case which mean existence only
two solutions, namely, $E_{\gamma} ({\bf u})$ and $E_{\gamma}^\ast ({-\bf u})$. $\Box$

Note, that from a practical point of view the first and second theorem cases do not really matter. Even a small change of the pupil aperture or non-separable pupil-plane wavefronts statistics which is common
resolve the separability issue. Thus, we will assume that the problem always has only two  conjugated wavefront solutions  $E_{\gamma} ({\bf u})$ and $E_{\gamma}^\ast ({-\bf u})$.  

As discussed above the conjugation sign of the solution can be established if it is known at an arbitrary point $\gamma_0$. This fact allows determining the conjugation state of the  measured wavefront by using either the asymmetry of the pupil aperture or introducing in the system an arbitrary defocus of known sign.
The image defocus not only provides the unique solution, but also improves the global convergence of the wavefront reconstruction algorithm, as discussed in Section~\ref{simulations}\ref{convergence}. Further, we  consider a pair of conjugated wavefront solutions as one solution, meaning that their conjugation state can be recognized if necessary. 
\begin{figure*}[t]
\begin{center}
\begin{tabular}{c}
\includegraphics[width=0.75\hsize]{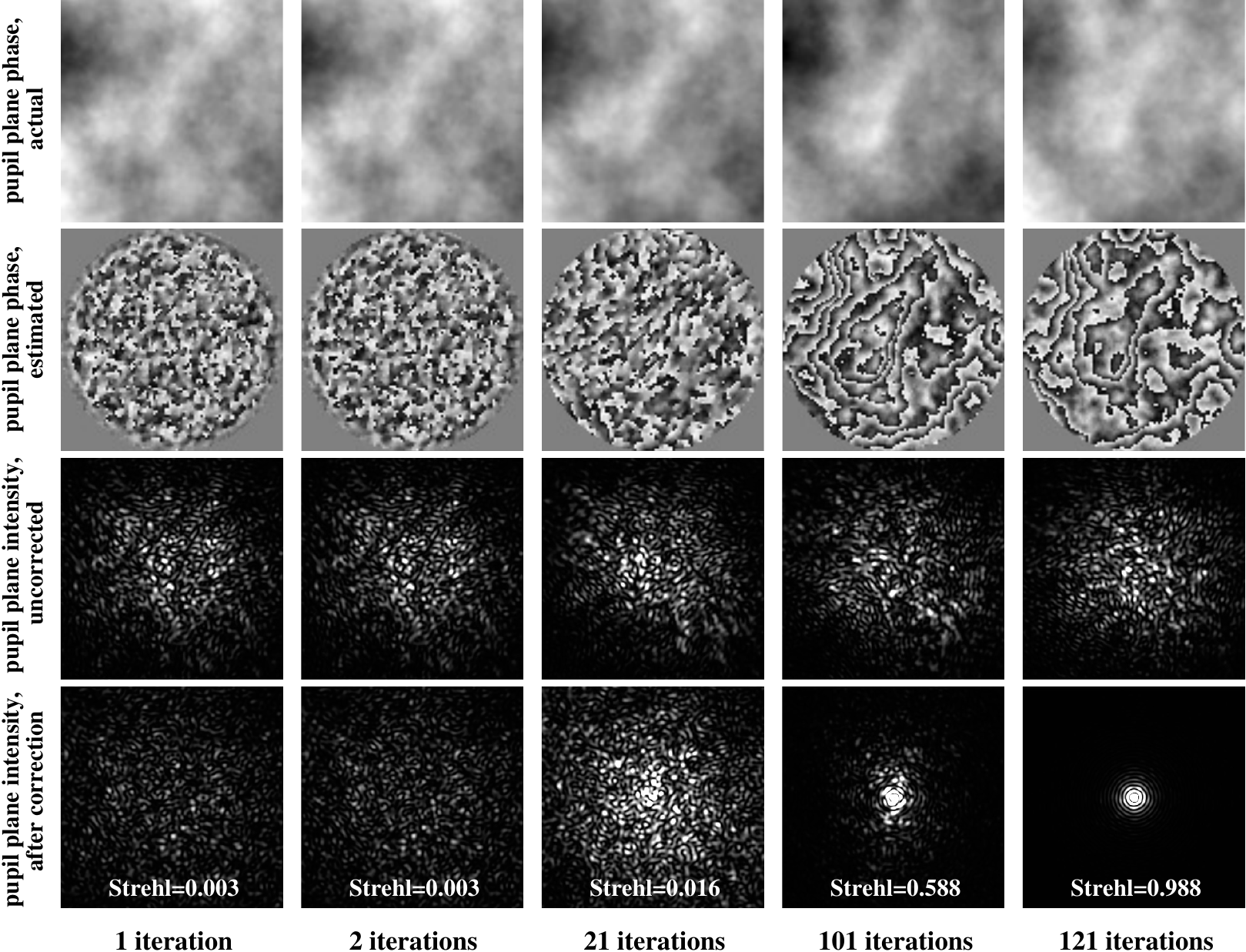}
\end{tabular}
\end{center}
\caption[fig2]
{\label{fig2}
{Restoration of a correlated sequence of random atmospheric disturbances 
(Kolmogorov turbulence, $D/r_0$=30, $\sigma_{\Delta\Phi}$=$\lambda/46$, $r_I$=0.99).
The image defocus parameter $a$ is  $\lambda/20$.
The actual and restored pupil-plane phases  are shown together with the  measured and corrected  focal-plane intensities obtained after 1, 2, 21, 101 and 121 iterations.}
}
\end{figure*}

\begin{figure}[t]
\begin{center}
\begin{tabular}{c}
\includegraphics[width=1.0\hsize]{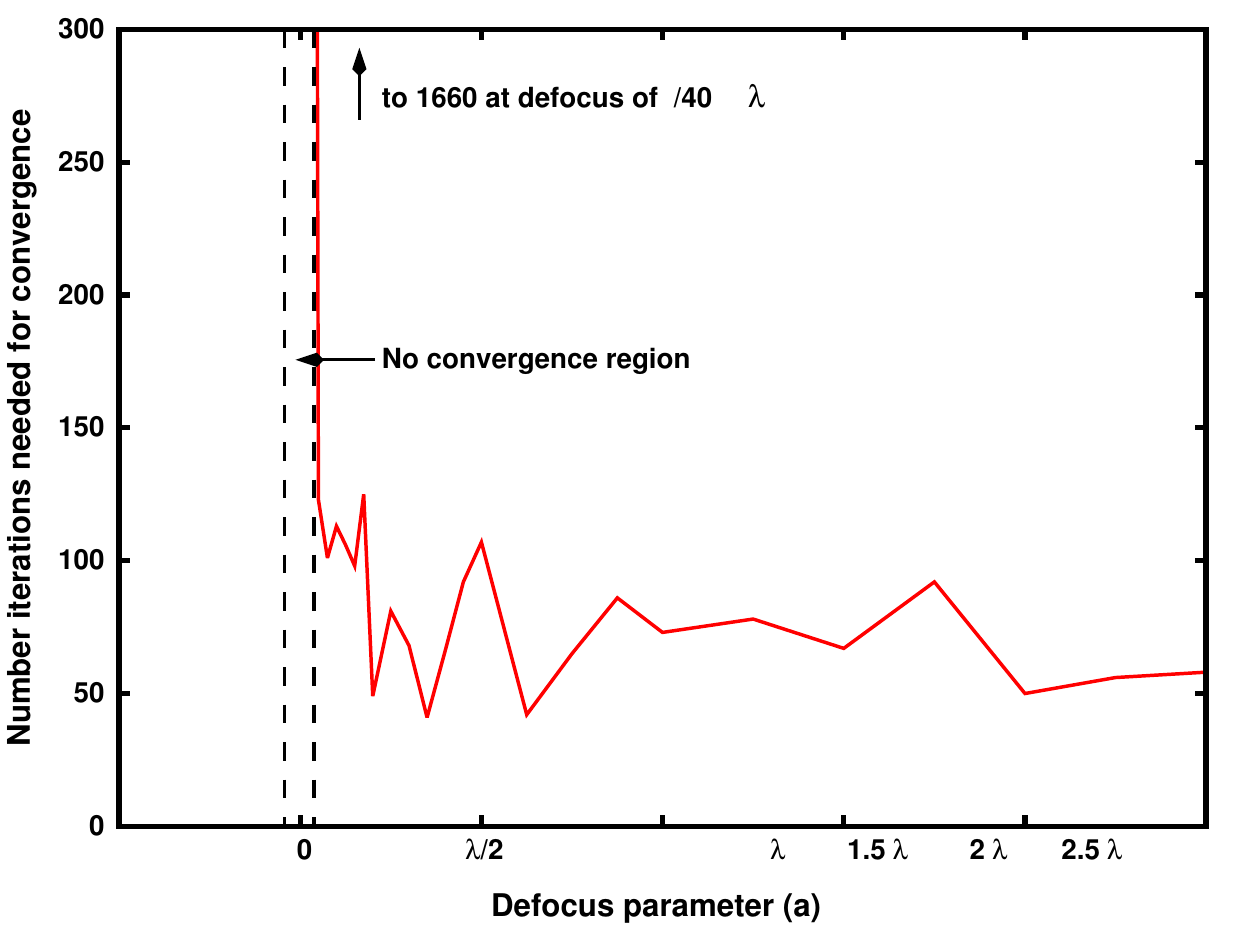}
\end{tabular}
\end{center}
\caption[fig3]
{\label{fig3} The algorithm convergence speed as a function of  focal plane image defocus (Kolmogorov turbulence, $D/r_0$=30,  $\sigma_{\Delta\Phi}$=$\lambda/46$, $r_I$=0.99).
}
\end{figure}

\begin{figure}[t]
\begin{center}
\begin{tabular}{c}
\includegraphics[width=1.0\hsize]{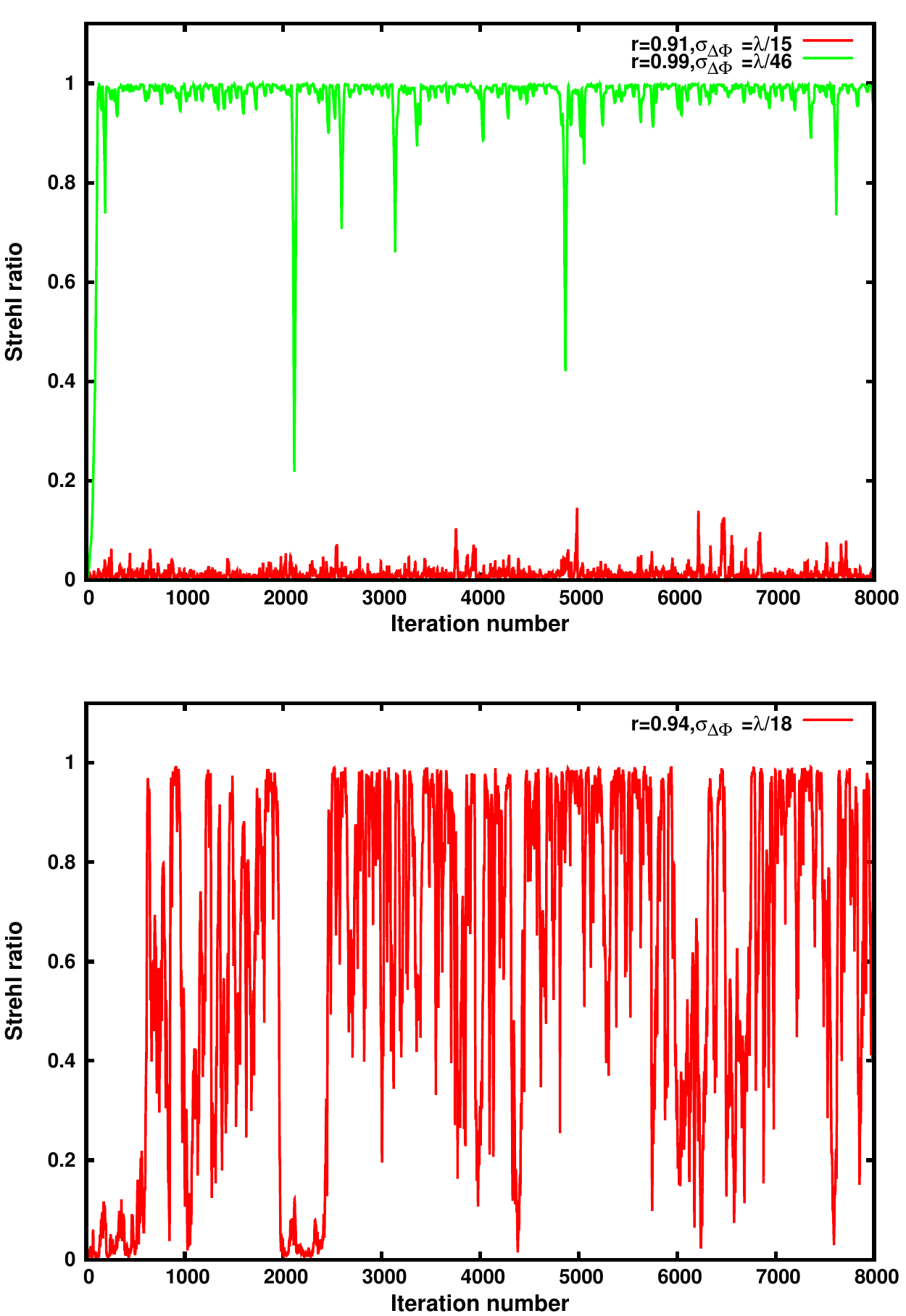}
\end{tabular}
\end{center}
\caption[fig5]
{\label{fig5}  Simulations of the algorithm convergence.  The Strehl ratio  $St$ as function of the iteration number is shown 
for $r_I$ equal to 0.91 and 0.99 (top),
and 0.94 (bottom) (Kolmogorov turbulence,
$D/r_0$=30, $a=\lambda/20$).
}
\end{figure}

\begin{figure}[t]
\begin{center}
\begin{tabular}{c}
\includegraphics[width=1.0\hsize]{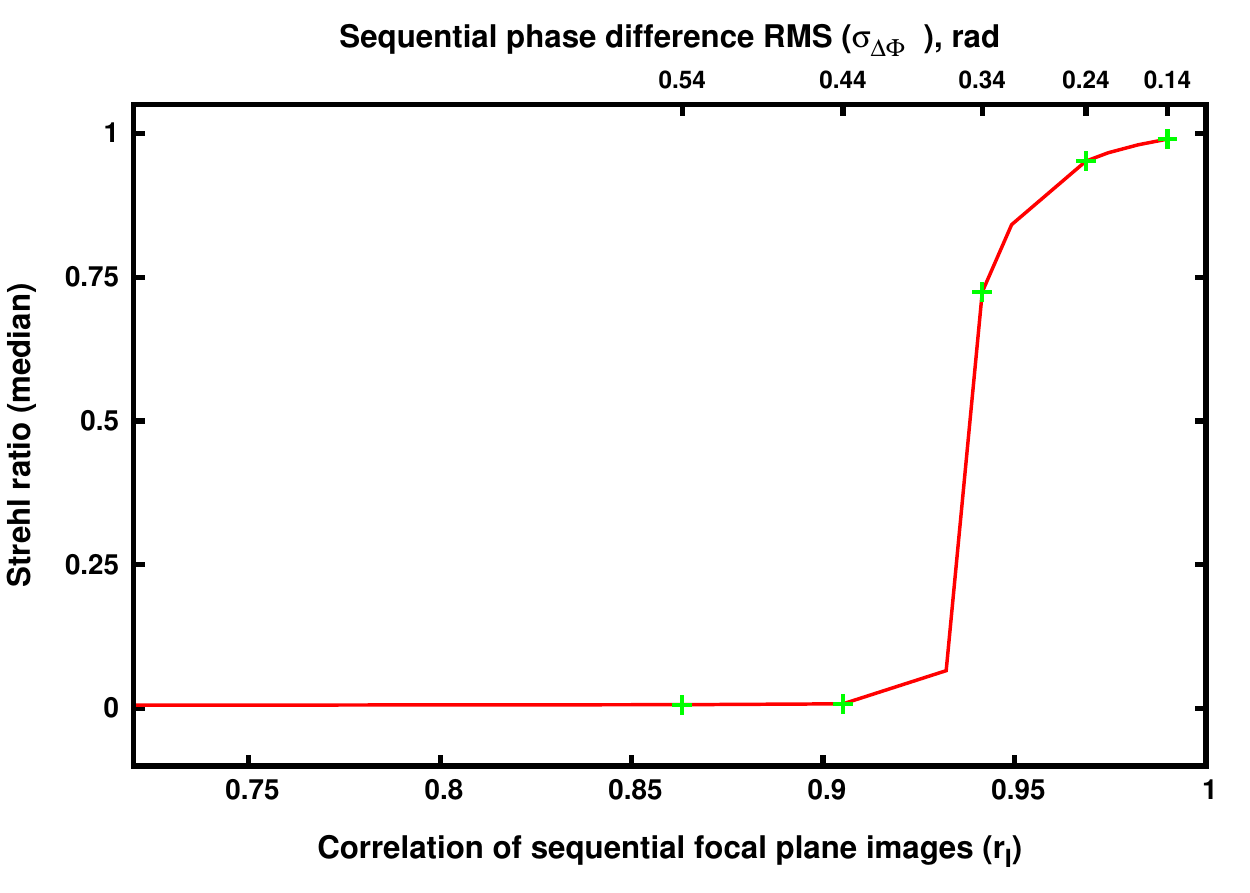}
\end{tabular}
\end{center}
\caption[fig4]
{\label{fig4}  The accuracy of the pupil-plane  phase restoration. The median value of Strehl ratios $St_m$  as function of the correlation coefficient $r_I$  and  the differential  phase  rms $\sigma_{\Delta\Phi}$ is shown
 (Kolmogorov ~ turbulence, $D/r_0$=30, $a=\lambda/20$).
}
\end{figure}

\section {Global convergence}
\label{methodc}

Existence of a unique wavefront solution does not mean 
convergence of the modified Gerchberg-Saxton algorithm  to this solution. To provide  global convergence of the algorithm, the following two conditions must be met:
\begin{description}
\item[\bf Condition 1:] {Successive focal-plane images must demonstrate sufficiently high correlation.}
\end{description}
{\bf Rationale: }Local convergence provided by the Gerchberg-Saxton algorithm must be preserved during the continuous phase transformation from the proposed modification. Hence, for any continuous pupil-plane phase change an iterative solution is always captured by the nearest local minimum or stagnation point \cite{Seldin_1990}. 
The local minima ``profile'' is being continuously deformed by the corresponding pupil-plane phase changes. Pupil-plane phase changes of sufficient magnitude can switch the iterative solution between different minima not connected by a continuous deformation. As a result, the algorithm may not necessarily be locally converging. 
However, in the case of sufficiently small sequential phase changes corresponding to the highly correlated successive focal-plane images, the captured iterative solution remains under the influence of a particular local minimum as long as needed for convergence.
Furthermore, the iterative solution remains trapped by the minimum while this minimum exists. 
To satisfy the local convergence condition, sequential focal-plane images should have negligibly small differences corresponding to high correlation between sequential pupil-plane phases.

\begin{description}
\item[\bf Condition 2:] In a sequence of focal-plane images, there must exist sets of completely uncorrelated images.
\end{description}
{\bf Rationale:} Local minima preclude convergence of the iterative solution to the global minimum. To provide global convergence, an efficient method to escape the vicinity of the local minima must exist. It should be taken into account that ambiguous multiple solutions produce identical focal-plane images. Hence, in accordance with Lemma 1, in a small area where an ambiguous solution creates a local minimum, differences in morphology of corresponding focal-plane images produced by the global and local solutions should be negligibly small.  However, the discrepancy between these solutions is growing as the pupil plane changes. At some moment, focal-plane intensities related to the global and local solutions become completely incompatible. Thus, the iterative solution can no longer be trapped by the previous local minimum and the iterative solution will enter the vicinity of another minimum (either global or local). The destruction of local minima is naturally associated with significant change in the morphology of focal-plane images. In a case where the temporal pupil-plane phase variations are strong enough to produce significant change in the morphology of focal-plane images only two outcomes are possible: (a) convergence to the global solution (the global minimum can be never destroyed) or (b) no convergence.

Numerical simulations and lab experiments related to the algorithm convergence problem (meaning global convergence) are discussed in Section~\ref{simulations}.  We only note here that there is a modification of the method where such convergence is always observed even in the presence of strong photon noise and background noise as described in Sections~\ref{simulations}\ref{ph_noise} and \ref{simulations}\ref{background_noise}.  

Finally, the proposed approach can be used to recover not only a dynamic pupil-plane wavefront, but can be also generalized for the case of static pupil-plane aberrations. In many cases correlated random pupil-plane phase
aberrations can be created as probes, for example, with a DM. The Gerchberg-Saxton loop can be used  to recover these probes combined with the static aberrations of the optical system. As soon as the iterative algorithm has found the global solution, gradual reduction of the amplitude of random DM probes forces the iterative solution to converge to the static aberration of the optical system.

\begin{figure}[t]
\begin{center}
\begin{tabular}{c}
\includegraphics[width=0.65\hsize]{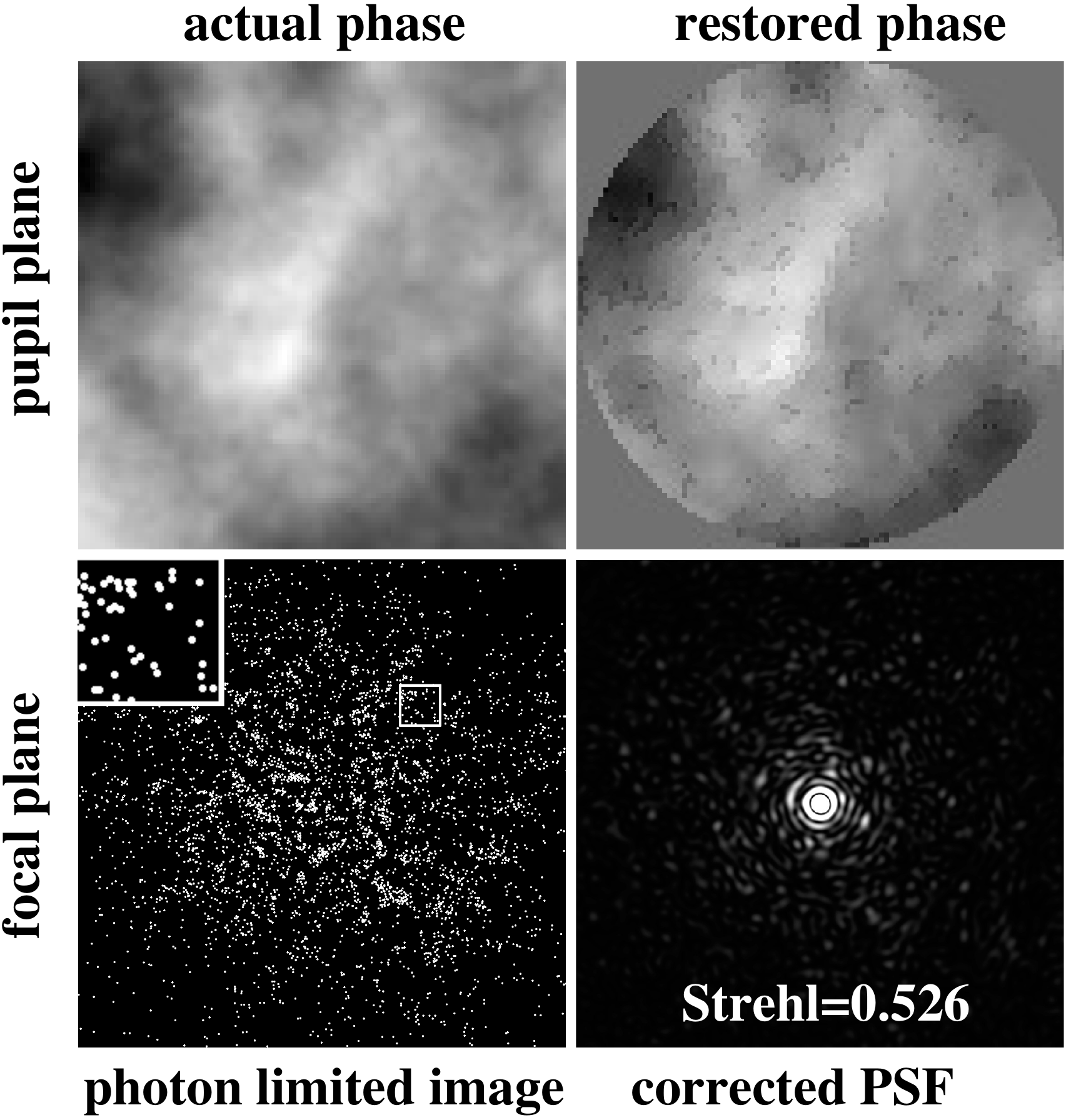}
\end{tabular}
\end{center}
\caption[fig7]
{\label{fig7} 
Restoration of a correlated sequence of random atmospheric disturbances 
(Kolmogorov turbulence, $D/r_0$=30, $r_I$=0.99, $a=\lambda/20$) in presence of photon noise. 
The actual pupil-plane phase and its unwrapped estimate after 105 iterations  are shown together with the
measured photon limited focal-plane image ($n_{ph}=2$ photons per speckle)  and the corrected PSF ($St$=0.526). The selected part of the photon limited image is shown with full resolution in the image upper left corner.}
\end{figure}

\section{Numerical simulations}
\label{simulations}

Conditions 1 and 2 discussed in the previous section are necessary but insufficient for algorithm convergence. For guaranteed convergence additional modifications of the method should be considered. From a practical point of view, it is also important to estimate the algorithm convergence speed and its noise sensitivity in cases where such convergence is provided.  
Numerical simulations described in this section address these questions.

 Although the statistics of the pupil-plane phase aberrations does not affect the solution uniqueness, it can influence algorithm convergence. Hence, we limited our simulations to two cases of practical importance. The first case (Model I) is the case of large dynamically changing phase aberrations caused by atmospheric turbulence.  We utilized an atmosphere model with spatial phase variations satisfying Kolmogorov statistics and with temporal variations described by a ``boiling'' turbulence model \cite{Pluzhnik_2004}. In accordance with this model, temporal variations are set by the inhomogeneities disintegration time $T_L$ that depends on the inhomogeneities size (scale) $L$ as $T_L\propto L^{2/3}$.
The chosen turbulence strength ($D/r_0$=30, $D$ is the entrance pupil size, and $r_0$
is Fried parameter) creates atmospheric wavefront disturbances typical for 
a 3~m telescope  under 1" seeing.
The temporal sampling for Model I is chosen such that the disintegration
time of the smallest spatial inhomogeneities is equal to the temporal sampling step (i.e.,  the smallest  inhomogeneities are completely uncorrelated for sequential wavefronts). The selected parameters  provide sequential focal-plane image correlation of 0.4-0.5  -- consistent with cross-correlation measurements of speckle-interferometric images at exposures comparable with  the atmospheric coherence time \cite{Ebersberger_1985, Vernin_1991}.

The second case (Model II) is the case of a moderate static pupil-plane aberration $\Phi_{stat}({\bf u})$. In this case, a sequence of uncorrelated phase screens with the spatial autocorrelation function 
$\propto\exp(-u/2\sigma^2_s)$ is used as random phase probes $\Phi_k({\bf u})$ ($k=1, 2,  3\dots$) that provide continuous transformation of the pupil-plane wavefront $\exp(i[\Phi_{stat}({\bf u})+\Phi_k({\bf u})])$. In our simulations, we chose the spatial correlation radius $\sigma_s$ equal to $D/12$ and the phase rms value of $\lambda/5$.

Unfortunately, neither Model I nor Model II data sets satisfy the local convergence condition, so a linear wavefront interpolation procedure is used to satisfy it.  In this procedure, pupil-plane phases $\Phi_{i*N}$  with sequence numbers $i*N$ are generated according to Model I or II. We refer to these pupil-plane phases as ``nodal probes''.
The gaps between the nodal probes are filled with phases
$\Phi_{k+i*N}=k*\Phi_{i*N}+(N-k)*\Phi_{(i+1)*N}/N$ where $i, k, N$ are natural numbers, and $k<N$. The correlation between two sequential pupil-plane phases can be arbitrarily set by choice of $N$, satisfying Condition 1. In our numerical simulations $N$ is set to 32 for Model I and 100 for Model II, which is enough to provide global convergence though does not guarantee optimal algorithm convergence speed.

To describe correlation between sequential focal-plane images and their corresponding pupil-plane wavefronts we use two parameters. One is the sequential focal-plane intensities correlation coefficient $r_I$ and the second is the sequential pupil-plane phase difference rms $\sigma_{\Delta\Phi}$. After each iteration, the aberrated wavefront can be corrected by applying the new wavefront estimate. The optical quality of the uncorrected and corrected point spread functions (PSF)
can be characterized by their corresponding Strehl ratios $St_I$ and $St$. To avoid wavefront comparison difficulties caused by wavefront unwrapping we use these parameters to evaluate algorithm performance. In some cases another evaluation metric we use is the median value of the Strehl ratios $St_m$ calculated for a sequence of corrected PSFs. The flat pupil-plane phase  is used as an initial pupil-plane phase guess in all simulations. 

\begin{figure}[t]
\begin{center}
\begin{tabular}{c}
\includegraphics[width=0.95\hsize]{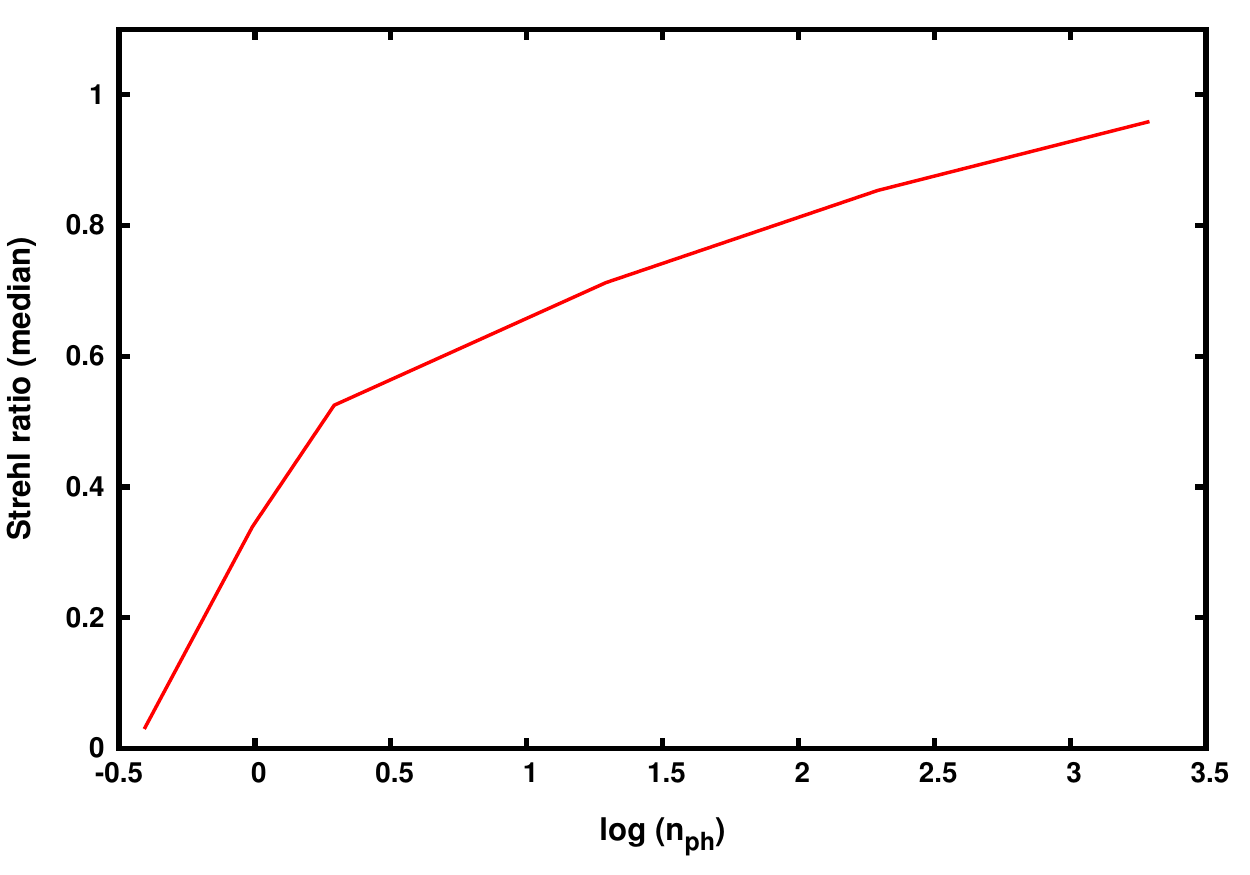}
\end{tabular}
\end{center}
\caption[fig8]
{\label{fig8}  
The accuracy of the  pupil-plane  phase restoration in the presence of photon noise. The median value of Strehl ratios $St_m$  as function of  number of photons per speckle $n_{ph}$ is shown (Kolmogorov turbulence,
$D/r_0$=30, $r_I$=0.99, $a=\lambda/20$).
}
\end{figure}

\begin{figure}[t]
\begin{center}
\begin{tabular}{c}
\includegraphics[width=1.0\hsize]{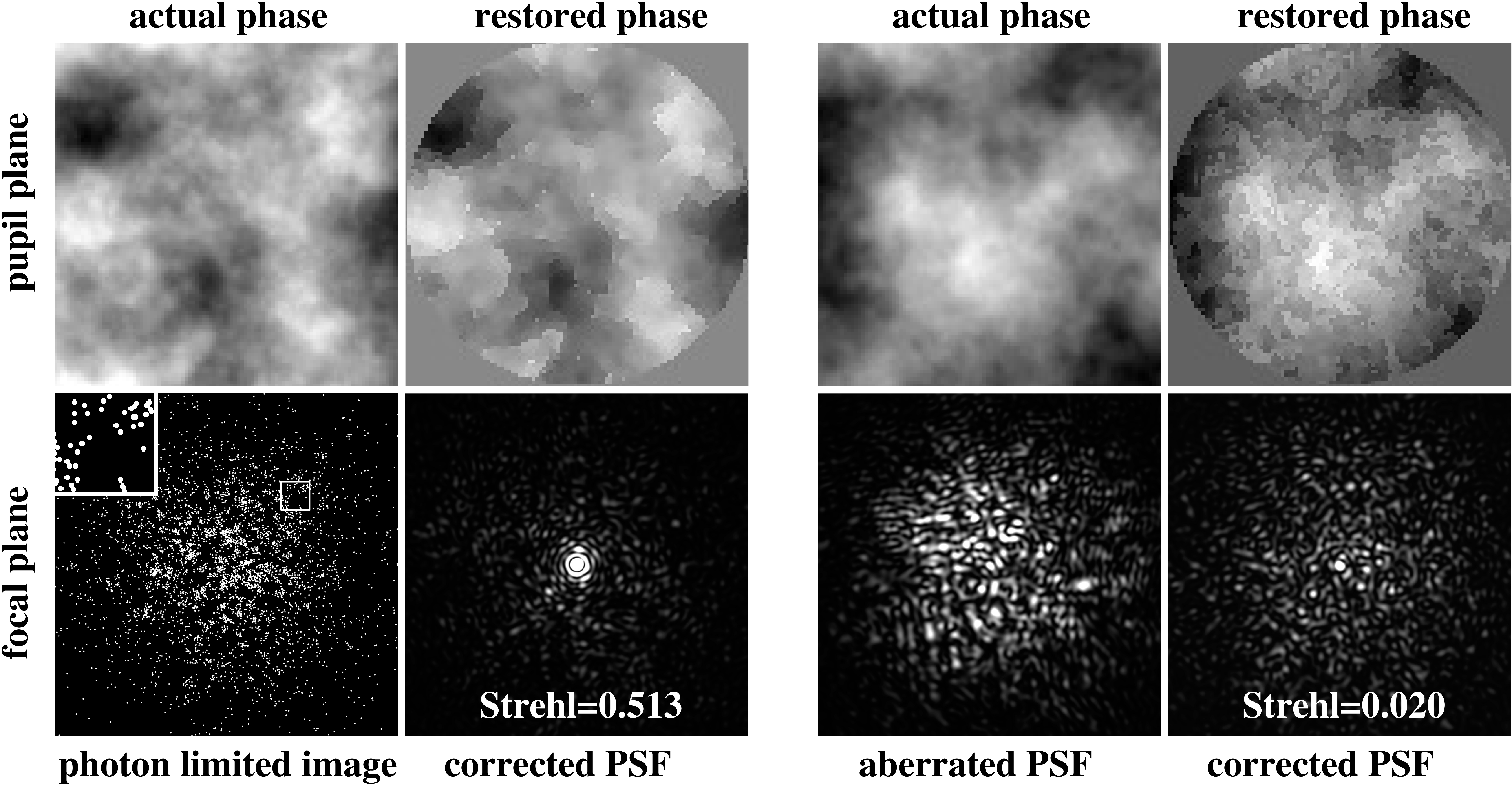}
\end{tabular}
\end{center}
\caption[fig9]
{\label{fig9}
Restoration of a correlated sequence of random atmospheric disturbances 
(Kolmogorov turbulence, $D/r_0$=30, $\sigma_{\Delta\Phi}=\lambda/11$, $r_I= 0.86$, $a=\lambda/20$) with  $r_I$ that  does not provide algorithm convergence  in the noiseless case. The restoration result after 130 iterations in the presence of photon noise ($n_{ph}=2$ photons/speckle, left) and the best restoration result in a noiseless run (300 iterations, right) are shown. All restored phases are unwrapped.
}
\end{figure}

\subsection{Algorithm convergence}
\label{convergence}
  
In the case of large phase aberrations
(similar to Model I) it is difficult to achieve algorithm convergence.
The absence of convergence can be explained by the wavefront degeneracy
presented in the focal-plane where the  convergence of the optical beam 
is replaced by the beam divergence. 

Each local or global solution forms an area in which iterative approximations are attracted by the solution. Due to the degeneracy, in a small region around the focus attraction areas formed by conjugated  wavefront solutions $E({\bf u},t)$ and $E^\ast ({-\bf u},t)$
overlap each other. The overlapping causes the appearance of saddle points in the vicinity of which algorithm convergence slows down. In the neighborhood of a saddle point the convergence speed is determined by the conditionality of 
the matrix $B_{jk}$ defined  in Section~\ref{theorem}. 
The matrix  conditionality rapidly decreases with increasing degrees of freedom
($\propto 2^K/K!$). The number of free  parameters $K$ describing a wavefront is a quantity of order of number of bright speckles observed in the PSF. This number is directly related to the pupil-plane phase rms. As aberrations become stronger the convergence rate can slow down and quickly enter into stagnation, resulting in algorithm convergence failure.  

To solve the convergence problem a combination
of the Gerchberg-Saxton modification with
the defocus based phase diversity is used. In this procedure each iteration includes three successive Gerchberg-Saxton steps which use  different focal-plane images. These three images are differently defocused by adding a small static defocus term  
$\Phi_{f}({\bf u})=a (8*u_x^2+8*u_y^2-D^2)/D^2$ to the pupil plane phase, where $a$ is  a defocus parameter. The first the image is in the focus.
Two other images have the defocus identical in modulus but opposite in sign.  
Sufficient defocus $a$ completely removes ambiguity caused by conjugated the solutions  $E({\bf u},t)$ and $E^\ast ({-\bf u},t)$  and recovers fast convergence in the vicinity of conjugated global solutions.
One example of such reconstruction performed for a Model I case is presented in (Fig.~\ref{fig2}). 
The example shows that a defocus as small as $\lambda/20$ is sufficient to ensure global convergence (including correct conjugation) after only 121 iterations, though without the defocus based phase diversity, convergence cannot be attained.

For smaller phase aberrations 
the iterative process converges even without the phase diversity use.
 However, in this case convergence
speed  is slow and
it is impossible to predict which of two conjugated 
solutions will be reached. 
As for the case of large aberrations the convergence rate
for small aberrations improves  
with suitable image defocus diversity. 
For example, the global solution for Model I with
$D/r_0=10$ can be obtained without image defocus
after 800-1000 Gerchberg-Saxton iterations. The same solution is achieved in just  40-60 iterations if the image defocus diversity  is used. Fig.~\ref{fig3} shows the dependence of the algorithm convergence speed on the defocus parameter for the case of large phase aberrations (Model I). The number of iterations needed for  algorithm convergence is determined as the number of the first iteration  for which $St>St_I$ and $St>St_m$. Global convergence of the algorithm is achieved as soon as the focal-plane degeneracy is removed by a sufficiently large defocus diversity.

The algorithm convergence is associated with sequential correlation of focal-plane images used for phase restoration.  In Fig.~\ref{fig5} the evolution of the phase retrieval quality $St$ is presented for three different cases simulated for Model I with the defocus parameter $a=\lambda/20$. 
In the first case $r_I=0.91$ and the algorithm does not demonstrate convergence because local convergence is broken by low sequential correlation. In the case with $r_I=0.99$ the iterative solution is easily captured by the global minimum and remains there for an indefinitely large number of iterations. In an intermediate case with $r_I=0.94$ there is a high chance for the iterative solution to find the global solution. However, due to continuity issues caused by insufficient sequential correlation 
the global solution cannot hold the iterative solution for a large number of iterations. 

Sequential correlation also affects retrieval accuracy. The average quality of the phase retrieval $St_m$
depends on the correlation coefficient $r_I$ as shown in 
Fig.~\ref{fig4}. The algorithm
does not converge until the correlation coefficient $r_I$ is less than 0.9 ($\sigma_{\Delta\Phi}>\lambda/15$). The algorithm converges only (meaning that it forms the Airy-like focal-plane  PSFs)  if $r_I$ is larger than 0.94 ($\sigma_{\Delta\Phi}<\lambda/18$).
Once $r_I$ reaches 0.97  ($\sigma_{\Delta\Phi}=\lambda/25$)
the Strehl ratio becomes larger than 0.95 for more than 50\%
of the corrected focal-plane images.
For $\sigma_{\Delta\Phi}=\lambda/46$ more than
90\% of the corrected images have $St$ larger than 0.95, though 5\% of them still show $St$ as low as 0.2  being, probably, affected by close local  solutions (Fig. \ref{fig5}, case with $r_I=0.99$). 
The algorithm convergence is restored immediately after destruction of local minima associated with these ambiguous solutions. 

The image defocus based phase diversity  and high sequential image correlation provide fast algorithm convergence to the global solution. In our simulations the global convergence is always observed in a few $N$ iteration cycles where $N$ is the number of correlated pupil-plane phase probes that fill the gap between two consecutive nodal probes.


\subsection{Photon noise}
\label{ph_noise}

Simulations have shown good performance in the noiseless cases considered above, however, in the presence of photon noise algorithm performance is expected to degrade in terms of algorithm convergence.  A surprising result from simulated case studies in which photon noise is included is that initial algorithm convergence speed actually improves in low flux cases. This convergence speed improvement comes at the cost of estimation accuracy as measured by an ultimately lower Strehl ratio.


A simulated case shown in Figure \ref{fig7} with photon flux as low as 2 photons per speckle (corresponding to approximately 5000 photons per frame) demonstrates convergence with a retrieved Strehl ratio $St$ of about $0.5$ after only 105 iterations. This convergence occurs from an initial wavefront with rms on the order of $\lambda$ and a corresponding $St$ of approximately $0.002$. 

To study the convergence of the wavefront retrieval method, the results of a set of simulated cases
obtained for atmospheric  wavefront disturbances (Model I) with sequential correlation $r_I=0.99$ 
are shown as a function of applied photon noise in Figure \ref{fig8}. $St$ strongly depends on  the number of photons per speckle.
Wavefront restoration accuracy improves with increasing flux, as expected, with a recovered Strehl ratio of $0.99$ for high photon flux levels. 

Operation under low light flux levels led to two unexpected benefits. One benefit is that for flux levels as low as 1-2 photons per speckle the modified Gerchberg-Saxton algorithm restores a moderate Strehl ratio $St$ even in cases where the sequential correlation is relatively low. This can provide algorithm convergence outperforming the ``no photon noise'' case. An example of such wavefront restoration is presented in Fig.~\ref{fig9}, where low flux has provided the algorithm convergence for as low $r_I$ value as $0.86$ after 130 iterations while a the ``no photon noise'' case shows no convergence.

\begin{figure}[t]
\begin{center}
\begin{tabular}{c}
\includegraphics[width=1.0\hsize]{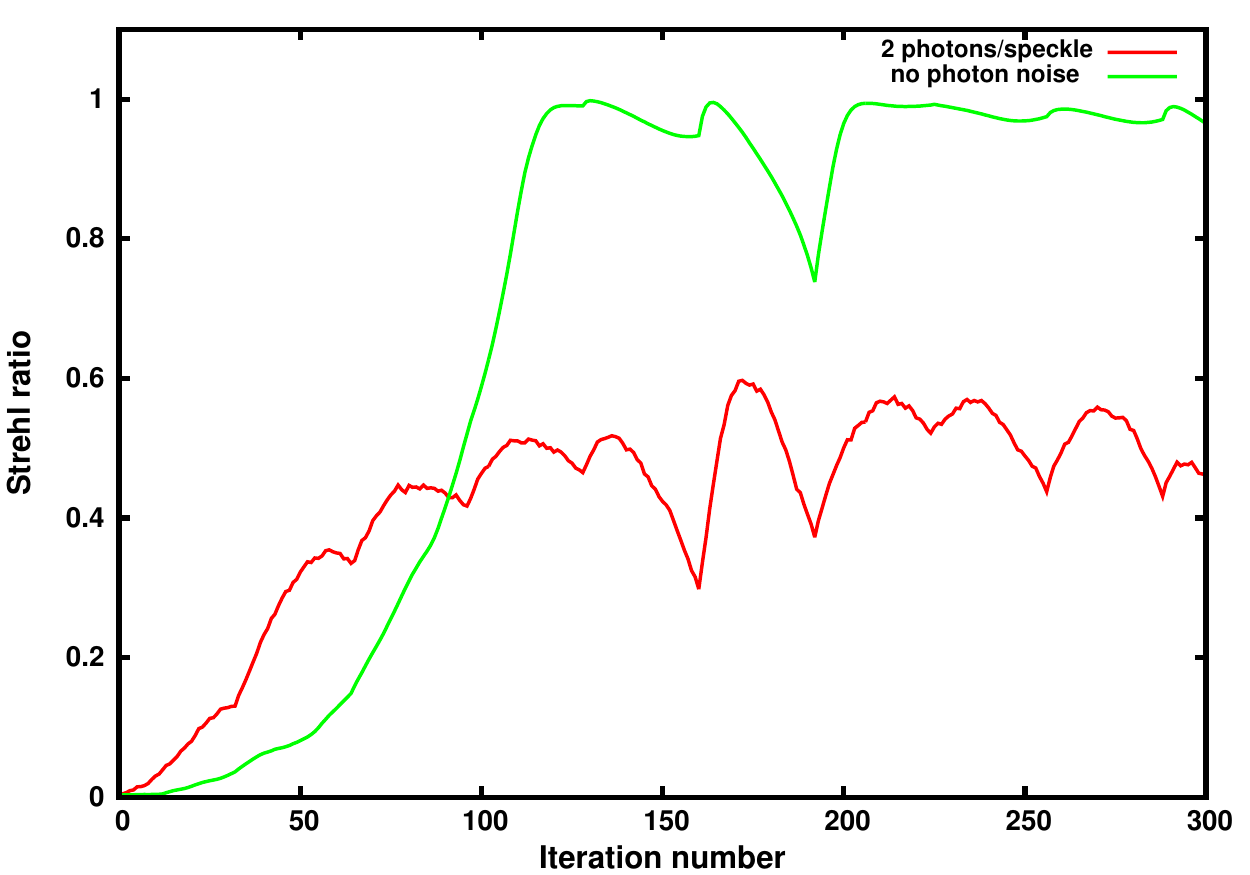}
\end{tabular}
\end{center}
\caption[fig10]
{\label{fig10} Comparison of algorithm convergence for restoration in both the absence and presence of photon noise. The pupil-plane phase restoration  quality $St$ as function of the iteration number is shown for the ``photon noise'' case
($n_{ph}\approx 2$) and for the noiseless case  running on the same data set that model Kolmogorov turbulence ($D/r_0$=30, $r_I$=0.99, $a=\lambda/20$). The ``photon noise''  related convergence curve is a result of data averaging obtained in 9 independent simulation runs. 
}
\end{figure}

 A second benefit is apparent from the $St$ dependence on the number of iterations in Fig.~\ref{fig10}. Algorithm convergence speed in the low photon rate mode is faster by a factor of 2-3 in comparison with ``no photon noise'' case. This faster convergence can be explained as follows. Assuming that a long exposure PSF can be roughly expressed as $\sim\exp[-r^2/2(\lambda/r_0)^2]$ and the sampling interval $\Delta x=\lambda/D$, number of sampling points $N_p$ needed for adequate pupil-plane phase description can be estimated as $(3D/r_0)^2\pi/4$. At the same time, the number of speckles in the focal plane $N_s=2.299(D/r_0)^2$ \cite{Roddie_1981} is about 3 times less than  $N_p$. At a photon rate of 1 photon per speckle,  
with high probability the photons will be detected in  the
centers of the brightest speckles.
The pupil-plane wavefront can be roughly presented as a superposition of spherical wavefronts whose centers coincide with the centers of detected photons but with relative phases unknown.  The number of unknown phases is  equal to the number of detected photons and is approximately equal to $N_s$.   As a result the photon limited phase retrieval problem has about 3 times fewer unknowns in comparison with  the ``no photon noise'' case. The reduced number of degrees of freedom simplifies the related optimization problem improving algorithm convergence but produces less accurate solutions. 

\begin{figure}[t]
\begin{center}
\begin{tabular}{c}
\includegraphics[width=1.0\hsize]{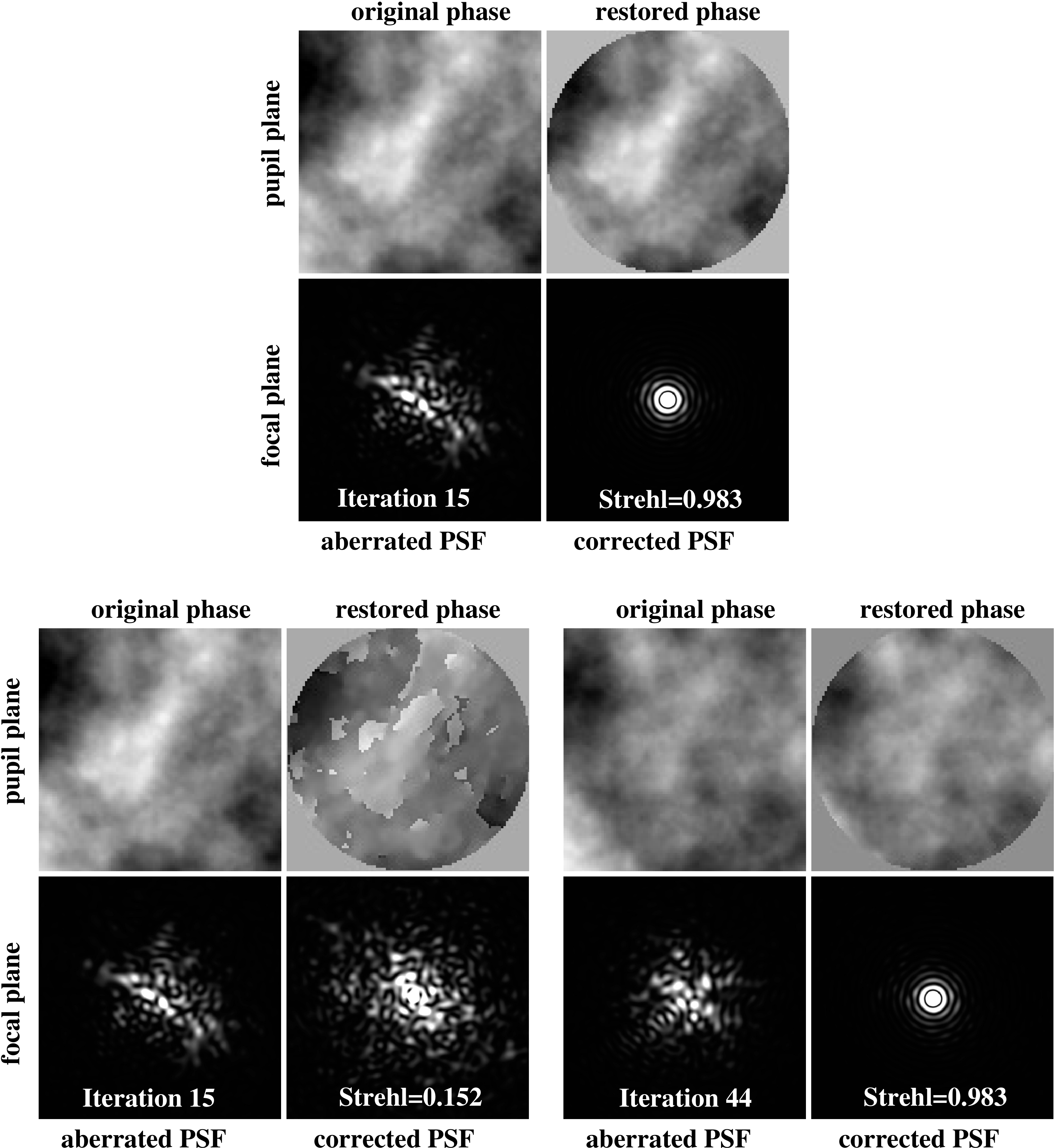}
\end{tabular}
\end{center}
\caption[fig11]
{\label{fig11}  The algorithm acceleration with artificial ``photonization'' procedure. Pupil-plane phase restoration results for runs with artificial  ``photonization'' (top, after 15 iterations, in first 10 iterations $n_{ph}=$ 2 photons/speckle) and without artificial ``photonization''
(bottom, after 15 and 44 iterations) are shown. In both cases the same correlated sequence of random atmospheric disturbances 
(Kolmogorov turbulence, $D/r_0$=10, $r_I$=0.99, $a=\lambda/20$) is used. All restored phases are unwrapped.}
\end{figure}

 Taking into account the observed convergence acceleration for the low photon rate case, a ``photonization'' algorithm modification can be formulated to improve convergence speed in the ``no photon noise'' case. This modification works as follows: the phase retrieval procedure  applies with low-photon rate images before switching to the ``no photon noise'' case in subsequent iterations. This procedure can be implemented by programmatic ``photonization'' of focal-plane images  with a photon noise generator. An example of this ``photonization'' procedure is presented in Fig.~\ref{fig11} where the algorithm with the ``photonization'' procedure converges to the global solution 3 times faster than the ``no photon noise'' case running on the same data set.

 Another phenomenon apparent in Fig.~\ref{fig10} is the periodic dipping in the $St$ curve with increasing iteration number. These dips are caused by the periodicity in the interpolation procedure used to generate correlated phase probes between uncorrelated nodal probes. Thus, every N iterations ($N=32$ for Model I) one of the old nodal probes is replaced with a new probe. Simultaneously, the local minima associated with the replaced probe are destroyed (as discussed in Section~\ref{methodc}) resulting in the observed periodic phenomenon as the iterative solution is released from its previous local minimum trap and continues convergence to the global minimum. There are moments when the iterative solution diverges to the global solution due to convergence to a new local minimum or due to slow convergence that does not allow the iterative solution to keep pace with the changing probes. The following nodal probe change starts the cycle again both releasing the new local minimum trap and adjusting convergence speed. Thus, the observed periodicity demonstrates the process of  rejection of local solutions. The periodical divergence can be deeper in the presence of photon noise which creates expected inconsistency between the global and iterative solutions.

\subsection{Phase retrieval with DM probes}
In previous sections we have shown how the Gerchberg-Saxton approach can be used to restore random, dynamically changing pupil-plane phases that satisfy Kolomogorov statistics. We will now show how the same approach works for different pupil-plane phase statistics described earlier  as Model II. The key requirement that provides algorithm convergence to the global solution is maintaining high sequential wavefront correlation. Thus, for random pupil-plane phases with rms asymptotically approaching 0, the Gerchberg-Saxton algorithm should also converge to 0 phase or equivalently to the static aberration of the optical system.

To simplify discussion we will not apply the defocus based phase diversity and ``photonization'' procedure improvements discussed above. Note that without these procedures, convergence speed of the algorithm is reduced and two conjugated phase solutions $\Phi({\bf u})$ and $-\Phi({-\bf u})$ may appear that will be transformed to the same orientation and polarity for comparison.

\begin{figure}[t]
\begin{center}
\begin{tabular}{c}
\includegraphics[width=1.0\hsize]{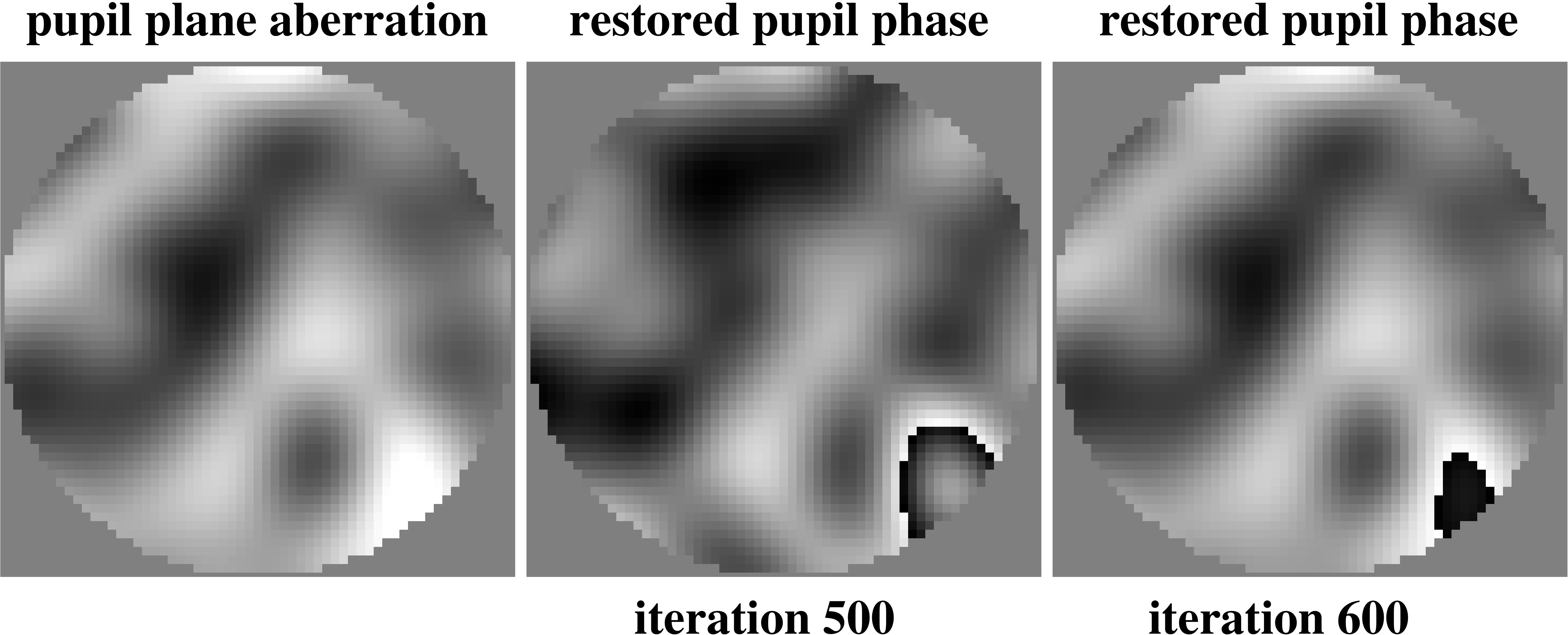}
\end{tabular}
\end{center}
\caption[fig12]
{\label{fig12} Restoration of a static optical aberration. The actual aberration (left), the restored pupil-plane phase after reaching the global solution (center) and the final pupil-plane phase solution (right) are shown. It took 500 iterations to find the global phase solution and 100 iterations more to estimate the actual pupil-plane aberration with rms of about $\lambda/150$.
}
\end{figure}

A numerical simulation case demonstrating restoration of a static pupil-plane wavefront is presented in Fig.~\ref{fig12}. The first sample of the Model II sequence uncorrelated with remaining samples plays the role of a static aberration  $\Phi_{stat}({\bf u})$ of the optical system. This static aberration is subsequently restored using a sequence of nodal probes and correlated Model II samples. The amplitude of phase  probes with rms is about $\lambda/5$ was not changing during the initial 500 iterations. After  catching  the global solution the amplitude of phase probes was reduced between successive iterations until it became negligible. In 100 iterations the amplitude of the phase probes was reduced from an rms level  $\lambda/5$ to $\lambda/5000$, which is below the phase restoration error of $\lambda/150$. The achieved phase retrieval accuracy is appropriate for most optical applications, and the approach can be used to determine static aberrations for any optical system where random, correlated probe phases can be created with a DM, for example . 

Unlike other phase probing methods, our solution does not rely on knowing what those probes are, so an accurate calibration of either the DM or the optical system is not a requirement. Additionally, geometrical calibrations needed to match pupil-plane and focal-plane wavefronts can be extracted from focal-plane intensity measurements and the procedure can be used as a calibration tool for other applications \cite{Giveon_2007, Giveon_2007_1, Krist_2015}. 

In addition to pupil-plane phase retrieval, we note that the random phase probe approach can also be used to restore pupil-plane amplitudes. An in-depth discussion of this application is, however, outside the scope of the current paper.

\begin{figure}[t]
\begin{center}
\begin{tabular}{c}
\includegraphics[width=0.8\hsize]{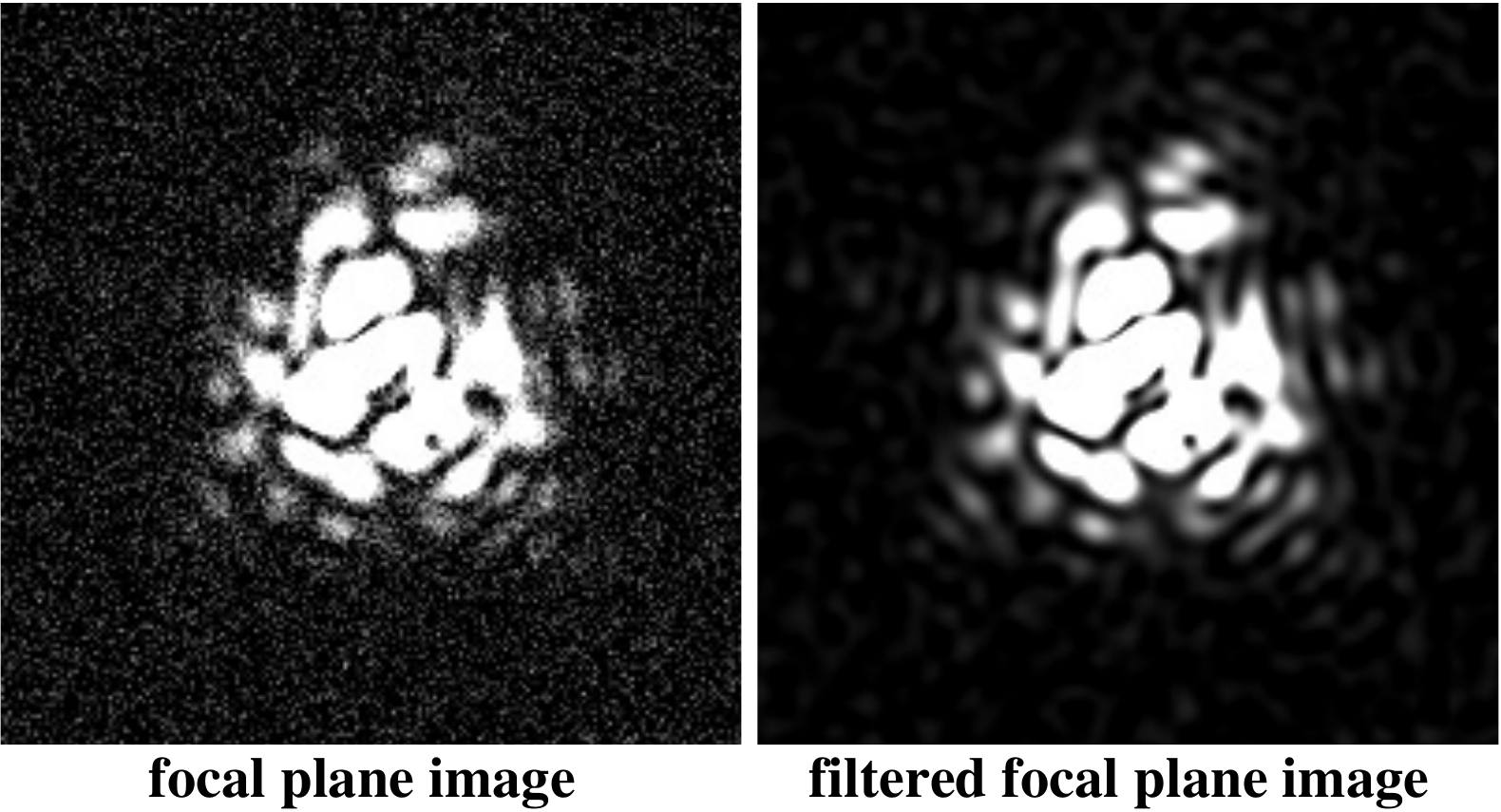}
\end{tabular}
\end{center}
\caption[fig14]
{\label{fig14}
Additive 1\% background noise in a simulated focal-plane image before (left) and after image filtering (right). The noise amplitude is measured relative to the average intensity of the brightest image speckles. This image also gives an estimate of the magnitude of aberrations needed to ensure the global algorithm convergence in the initial
stage of our laboratory demonstration (Section~\ref{experiment}).
}
\end{figure}

\begin{figure}[t]
\begin{center}
\begin{tabular}{c}
\includegraphics[width=0.9\hsize]{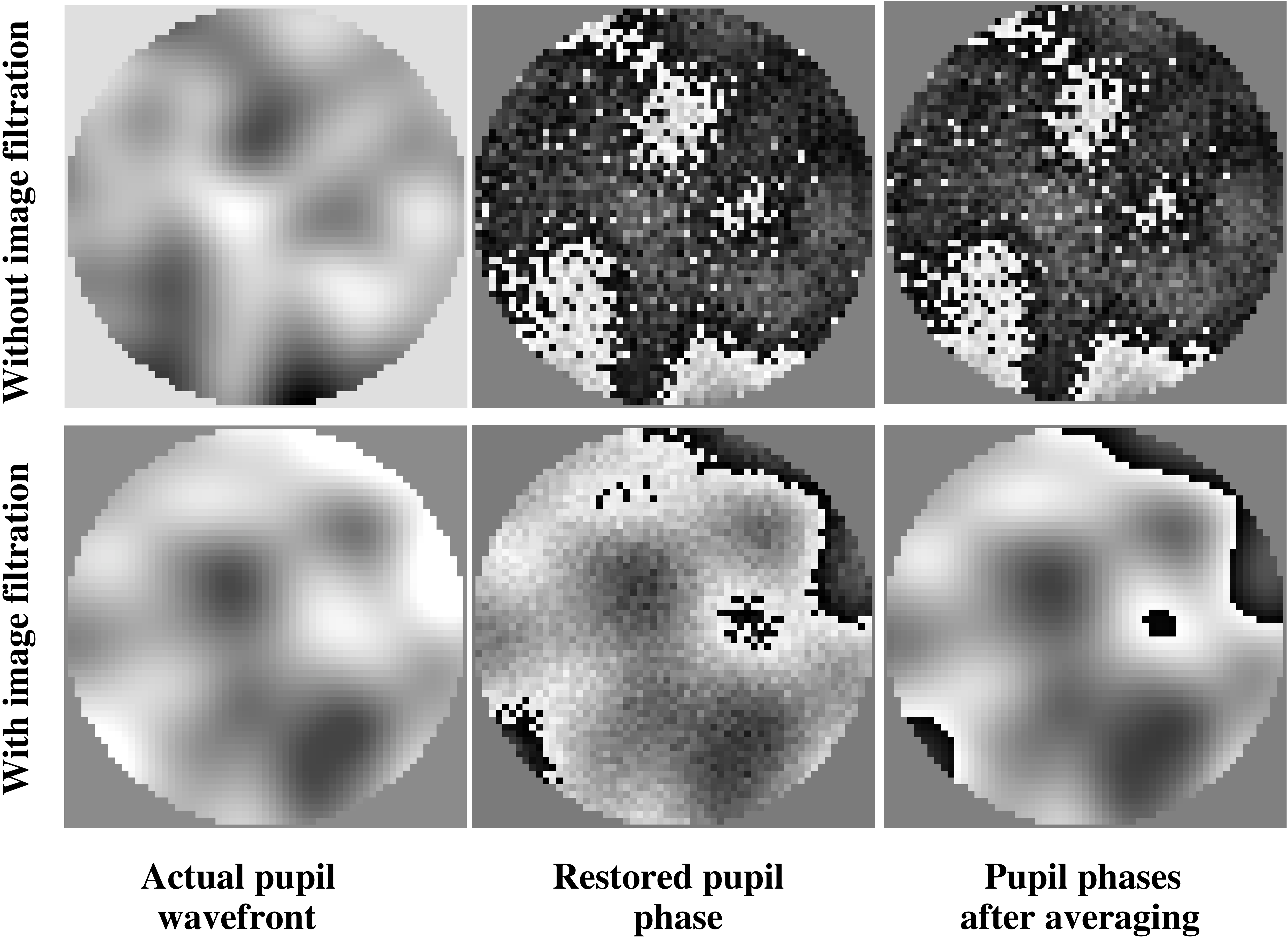}
\end{tabular}
\end{center}
\caption[fig15]
{\label{fig15}
The additive background noise averaging in the pupil-plane phase reconstruction  procedure. The actual (left) and restored  pupil-plane phases without (center) and with  (right) background noise averaging are shown  for two different cases:  the phase retrieval  without (top) and with  focal-plane image  filtering (bottom).
}
\end{figure}

\subsection{Background noise} 
\label{background_noise}

 Wavefront retrieval can also be affected by background image noise such as focal-plane detector read noise. An example of focal-plane image with additive $1\%$ rms background noise is shown in the left panel of Fig.~\ref{fig14}. 

Results of wavefront retrieval simulations for a static pupil-plane aberrations with $1\%$ background noise are presented in Fig.~\ref{fig15}. The background noise results in  slower convergence compared with the ``no noise'' case and also causes white noise that degrades the pupil-plane phase solution as clearly seen in the top, center panel of Fig.~\ref{fig15}. An averaging procedure can be formulated to improve the accuracy of the phase estimate and mitigate the induced white noise:

\begin{figure*}[t]
\begin{center}
\begin{tabular}{c}
\includegraphics[width=0.67\hsize]{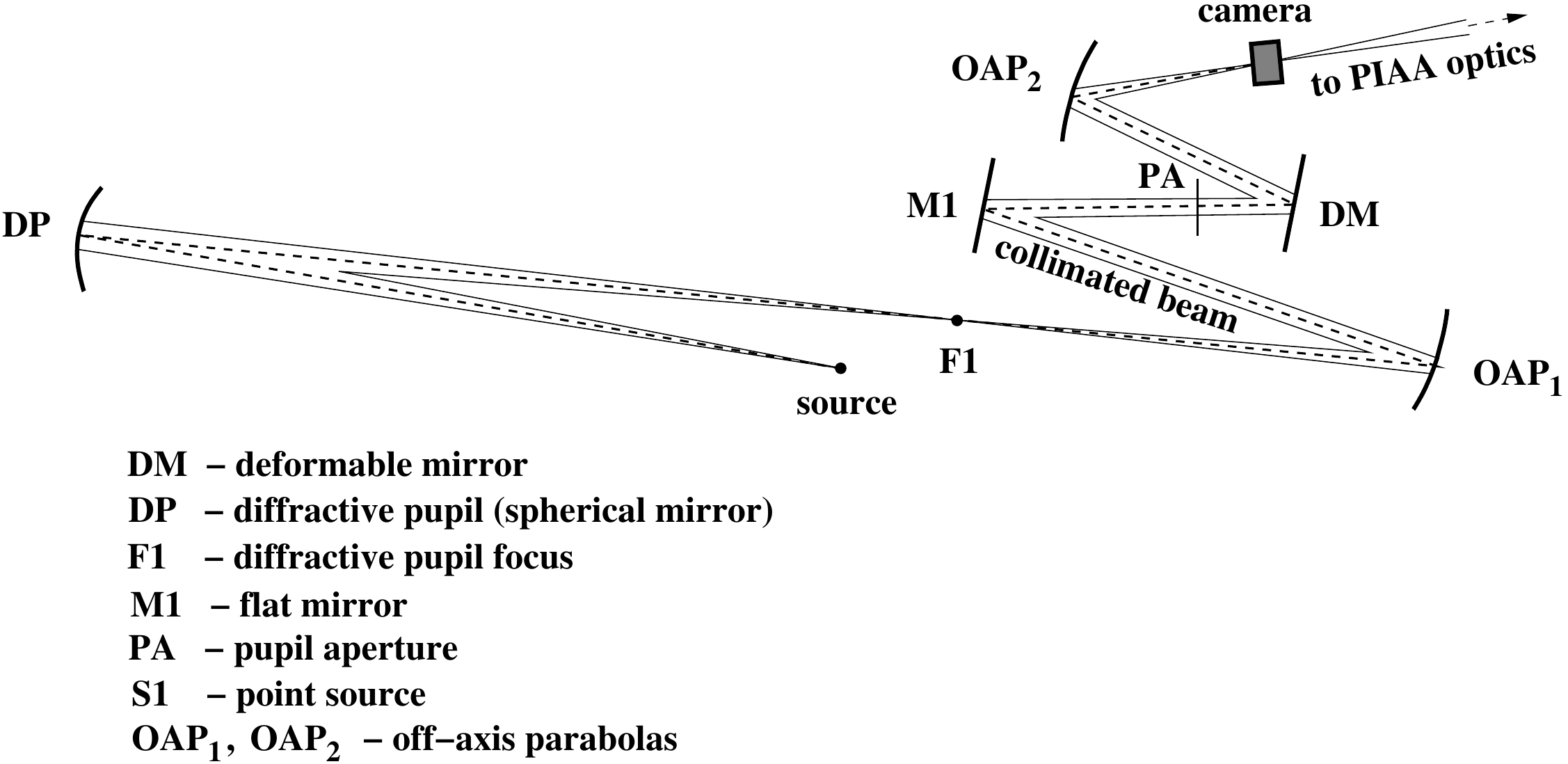}
\end{tabular}
\end{center}
\caption[fig16]
{\label{fig16}
The experiment layout.
}
\end{figure*}

\begin{equation}
E({\bf u})=|E({\bf u})|\exp[i\Phi({\bf u})]=\langle E_i({\bf u})\rangle.
\label{eq20}
\end{equation}
In Eq.~\ref{eq20} the final pupil-plane phase estimate $\Phi({\bf u})=\arg[\langle E_i({\bf u})\rangle]$, and $\langle...\rangle$ represent averaged sequential wavefront estimates $E_i({\bf u})$ obtained after the wavefront retrieval procedure has converged to the static optical aberration.

Although the averaging procedure can improve the final pupil-plane  wavefront estimate, it requires simple image filtering to be applied to every image used in reconstruction (Fig.~\ref{fig15}).  Each image should be low-pass filtered in the frequency domain by removing all frequencies beyond the pupil cut-off frequency as shown in the right panel of Fig.~\ref{fig14}.  This simple filtering procedure shows significant improvements in wavefront reconstruction quality in the presence of background noise with Strehl ratio increasing from 0.68 (without filtering) to 0.91 (after filtering). When the averaging procedure is also applied to the filtered reconstruction, a final Strehl ratio of 0.998 is obtained after  averaging  of 1000 wavefront estimates. 


\section{Experimental results}
\label{experiment}

The main results of the  numerical simulations have been  confirmed in a set of laboratory experiments performed at the NASA Ames Coronagraph Testbed \cite{Belikov_2010}. During these experiments the phase in the entrance  pupil of the PIAA (Phase Induced Amplitude Apodization) coronagraph \cite{Guyon_2005} was restored using a set of random phase probes created by the DM located in the entrance pupil.

\subsection{Experiment description}

Except for a few minor differences the PIAA coronagraph setup at NASA Ames is similar to the setup used in our EXCEDE (EXoplanetary Circumstellar Environments and Disk Explorer) demonstration \cite{Sirbu_2016}.The front-end of this setup  that feeds into the PIAA coronagraph was used in the wavefront retrieval experiments.  The experiment optical layout is shown in Fig.~\ref{fig16}.  The spherical wave formed by the  point source S1 (655nm laser) passes through the diffractive pupil (DP) \cite{Bendek_2012} and is focused in the front-end focus F1 at a distance of 531 mm from the diffractive pupil. The distance between the point source and the diffractive pupil is 473 mm. The diffractive pupil is a spherical mirror with curvature radius of 500 mm and features a low frequency diffractive grating on its surface. The beam is collimated farther by an off-axis parabolic mirror OAP1, and is reflected by the fold mirror M1 and the deformable mirror DM.   Finally, the collimated beam is focused by the second off-axis parabola OAP2 in the first focus of the PIAA coronagraph. The focal lengths of OAP1 and OAP2 are 305 mm and 127 mm respectively. The diffractive pupil is optically conjugated with the DM and both are optically conjugated with the first PIAA mirror. A circular pupil stop PA with the diameter of 9.3 mm is used to restrict the beam size. The pupil stop is located upstream of the DM at the minimal distance that prevents beam vignetting.


The focal-plane images are focused on the Basler acA3800-14um camera with a pixel size of 1.67 $\mu$m that is small enough to provide an appropriate image sampling (the system $\lambda f/D$=9.2~$\mu$m). The camera was operated in video mode with 14 twelve-bit frames/sec. After averaging 100 frames, each 600$\times$600 raw focal-plane image was clipped down to 512$\times$512 before usage in the wavefront retrieval procedure. Frame averaging was needed to reduce camera background noise and increase dynamic range.


Our experiments used a Boston Micromachines MEMS (Micro-Electro-Mechanical) DM featuring $32 \times 32$ actuators. Image sampling set a spatial resolution of 86 pixels across the pupil ($\approx$2.5 pixels/DM actuator) as measured from the image power spectrum.

The fiducial pattern applied to the DM was the flat DM surface with three actuators poked in the same ``positive'' direction and one actuator poked in the opposite direction. The related DM voltage map \cite{Gavel_2009} and focal-plane image are presented in Fig.~\ref{fig17}. The maximal amplitude of poked actuators reaches 140 nm resulting in no more than 0.02 degradation in Strehl ratio (in comparison with the flat DM case).

The wavefront retrieval algorithm was run as follows. First, a set of independent nodal probes was generated similar to Model II with a correlation radius equal to 12 DM inter-actuator distance. The corresponding DM voltages were calculated based on a quadratic model of the DM deflection curve with amplitude normalization chosen experimentally to be sufficiently strong such that the main lobe of the focal-plane image (Fig.~\ref{fig17}) was destroyed (Fig.~\ref{fig14}). Second, the iterative retrieval algorithm was run with $N$ partially correlated phase probes between the nodal probes. For the first 600 iterations, $N$ was set to 200 and the amplitude of  probes  was fixed until convergence to the global phase solution. Subsequently, the phase  amplitude of the nodal probes  was exponentially reduced with the successive amplitude ratio of 0.9 until the probes became negligible after 2000 iterations (in total).
To speed-up convergence during this step, $N$ was gradually reduced to 100, 70, and 50. Third, the final phase solution was obtained by applying the averaging procedure to 1000 successive wavefront estimates obtained with negligible phase probes.

\subsection{Phase retrieval results} 

\begin{figure}[t]
\begin{center}
\begin{tabular}{c}
\includegraphics[width=0.9\hsize]{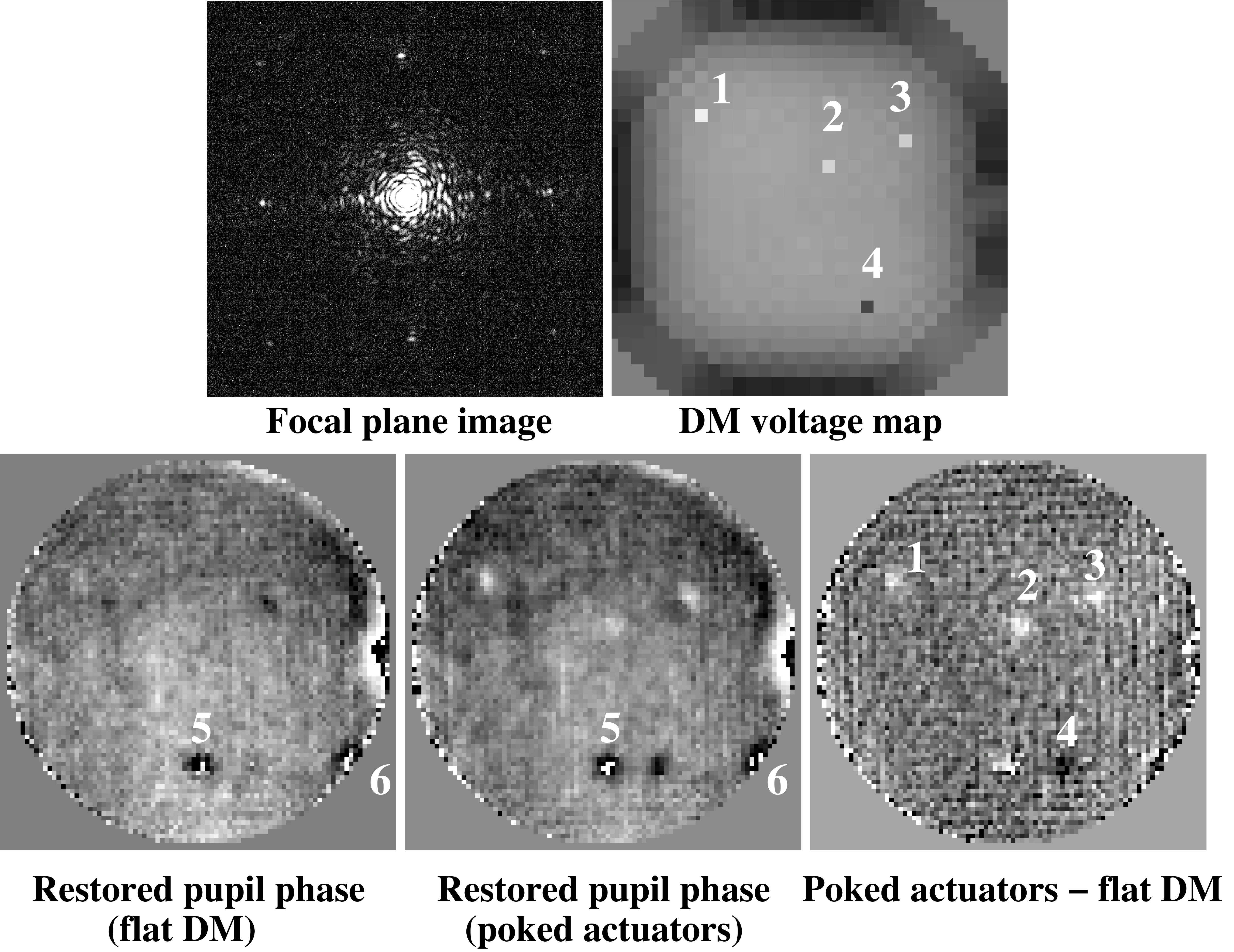}
\end{tabular}
\end{center}
\caption[fig17]
{\label{fig17}
The laboratory experiment results. The restored pupil-plane phase  is shown  together
with a focal-plane image and the DM applied voltage. The fiducial poked actuators (1, 2, 3, and 4)
and  malfunctioning DM actuators (5 and 6) are marked. Actuators 1, 2 and 3 have positive phase while the malfunctioning actuators and actuator 4 have negative phase. 
The restored flat DM phase (unaveraged) and the difference between the 
poked actuators phase map and flat DM  phase are also shown.
}
\end{figure}

The restored pupil-plane phase is presented in Fig.~\ref{fig17}. Two separate wavefront retrieval run results are shown, one using a flat DM and a second using the fiducial DM pattern. Since the DM is conjugate to the entrance pupil, DM feature details can be clearly seen in the estimated phase results. First, the poked actuators are recognizable in locations and with polarities matching the applied fiducial pattern. Recovered actuator poke amplitudes match the commanded signal (see discussion in Section~\ref{experiment}\ref{reliability}). Ignoring small tip/tilt and defocus terms, the retrieved phase map appears flat except across the right edge of the pupil where a phase growth of $0.7 \lambda$ is observed and two (known a priori) malfunctioning actuators marked as ``5''  and ``6''. 
There are also a few DM areas with weak phase depression
(in relation to the flat DM map) 
likely caused by DM flattening errors or aberrations produced
by other optical elements.
Finally, there is a poorly recognizable grid-like  structure that is likely caused by the quilting pattern formed by DM actuators \cite{Sirbu_2016_1}.

All the above features, except the DM quilting pattern, are highly repeatable and consistent between different wavefront retrieval runs. The periodic DM quilting pattern cannot be firmly recognized because the spatial resolution of the retrieved pupil-plane phase is insufficient to resolve it.  The phase map resolution across the pupil plane is limited by two main factors: (1) sampling and (2) dynamic range of detected focal-plane images.
To allow restoration of the amplitude and location of the DM grating the camera dynamic range should be sufficient to linearly detect  weak focal-plane speckles with contrast of a few times $10^{-5}$ near the diffractive peaks produced by the DM grating  (contrast of $10^{-3}$).  These speckles interfere with the diffractive orders and are the most sensitive to the spatial shift of the grating.
To expand the dynamic range  a high dynamic imaging tool was implemented that allows to combine a set of images taken with different exposure times in one image with dynamic range spanning 5 orders of magnitude. We also doubled pupil-plane spatial resolution by increasing the size of collected images to 1200$\times$1200 and clipping them down to 1024$\times$1024. 

\begin{figure}[t]
\begin{center}
\begin{tabular}{c}
\includegraphics[width=0.9\hsize]{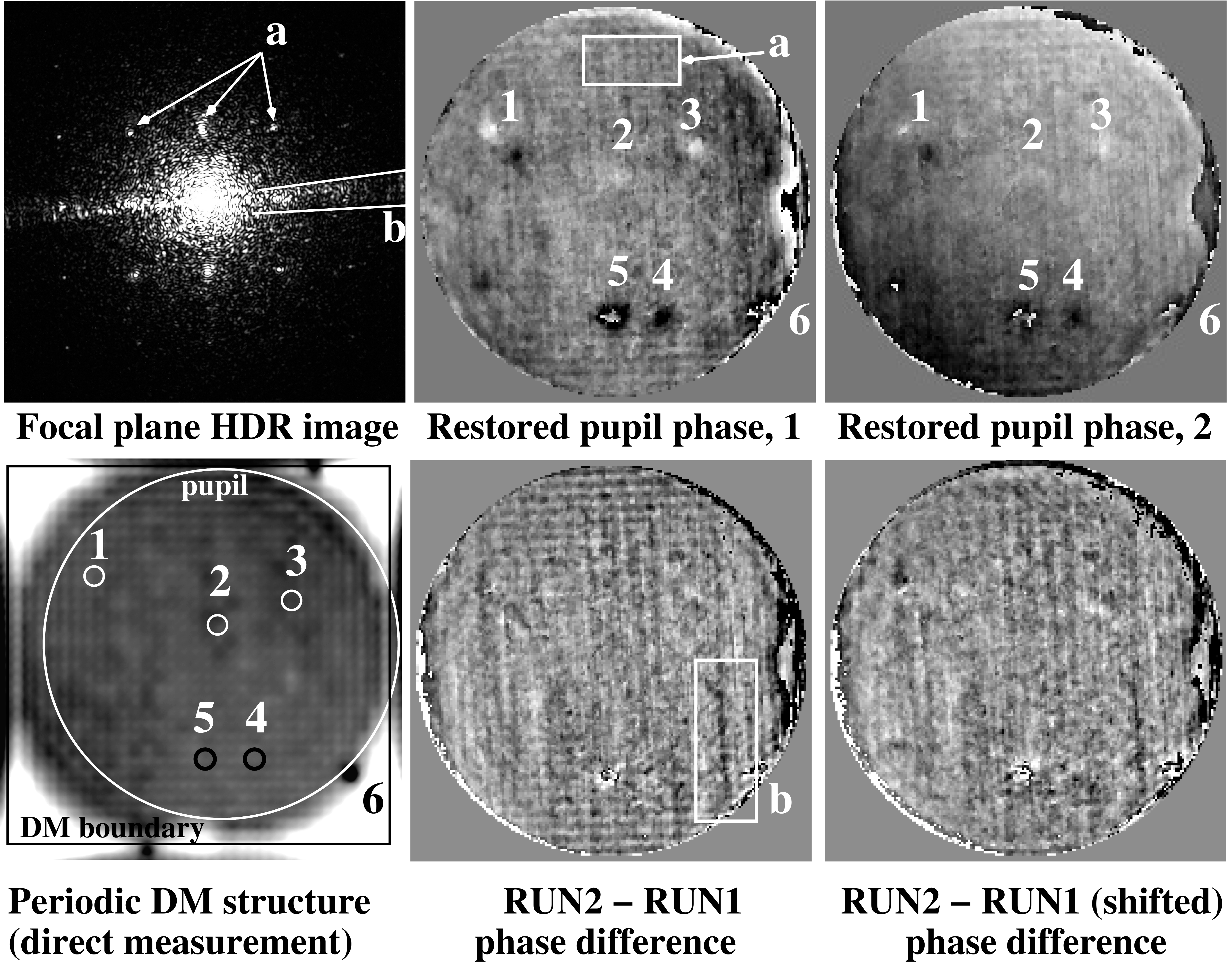}
\end{tabular}
\end{center}
\caption[fig18]
{\label{fig18}
 The pupil-plane phase restoration with implemented  high  dynamic range (HDR) imaging procedure. The restored  phase maps  obtained in two  independent
phase retrieval runs are shown. The phase difference between these two maps without lateral shift and with 
1 pixel of horizontal and 3 pixels of vertical
relative shift are also shown.
The direct interferometric measurement of the flat DM
surface is presented for comparison \cite{Gavel_2009}. 
The following DM produced features are marked:
poked actuators (1, 2, 3 and 4), malfunctioning actuators (5 and 6)
the DM grid related  structure  (a),
vertical structure associated with OAPs diamond turning errors, the DM boundary
and the pupil aperture. 
}
\end{figure}

Better sampling and imaging dynamic range improve the resolution of the phase solution. Two independent pupil-plane phase estimates with increased resolution of 5 pixels/actuator (172 pixels across the pupil)  are shown in Fig.~\ref{fig18}. We analyze these phase maps for consistency in Section~\ref{experiment}\ref{reliability}.


\subsection{Phase retrieval reliability}
\label{reliability}

In comparison with the  2.5 pixels/actuator  resolution map, the DM surface grating with amplitude of about $\lambda/30$ nm ($\sim$20~nm) is clearly visible in higher resolution maps. However, a more detailed analysis indicates some systematic differences  in the restored phases.
 Following is an analysis of  the observed differences:

(1) Low order modes of the two estimated phase maps are practically identical, except for tip/tilt differences between different trials. Since we did not set a certain focal-plane image position in the frame during the phase restoration, the observed tip/tilt simply reflects  the changes of the source location caused by the long term instability of the optical setup.

(2) Fiducial actuator pokes are well-matched both in location and amplitude being completely subtracted in the phase difference map (Fig.~\ref{fig18}). Relative amplitudes of poked actuators (Table~\ref{table1}) are consistent with amplitude ratios derived from  direct DM deflection measurements \cite{Gavel_2009}. The absolute poke amplitudes
are calculated based on the quadratic DM deflection  model  with the maximal DM stroke estimated   as the average of the known deflection based value  
and the inter-actuator maximal stroke from \cite{Morzinski_2008}. 

(3) A systematic pattern is seen in the phase difference map (Fig.~\ref{fig18}) that consists of two components. The first component is a regular grid-like structure produced by the DM surface grating. Setting a 1 pixel horizontal shift and a 3 pixel vertical shift in one of the phase maps results in the DM grid artifact disappearing when subtracted. This strongly suggests an ambiguity in the spatial location of the grid pattern.

(4)  The second systematic component appears as irregular, mainly vertical stripes across the pupil.  These
strips are, probably, introduced by the optics and variability in the stripes position can be explained  by slow drift in the optical setup. The source of this structure may be  errors in the surface of OAPs produced by the diamond turning process. These diamond turning  errors are detectable in reflected light at large incidence angles and are also responsible for the tilted horizontal band
visible in the focal-plane PSF (marked with ``b'' in Fig.~\ref{fig18}). 

 (5) The phase difference  rms of $\lambda/14$ is dominated by the discussed systematic pattern.

\begin{table}[t]
\begin{center}
\begin{tabular}{|c|c|c|c|c|}
\hline
Poked& \multicolumn{2}{|c|}{Actuator  amplitude}& \multicolumn{2}{|c|}{Relative amplitude}\\
\cline{2-5}
actuator&restored&applied&restored&applied \\
\hline
1&~135~nm  &~173~nm&~1.0 &~1.0\\
2&~75~nm   &~102~nm&~0.56&~0.59\\
3&~94~nm   &~133~nm&~0.70&~0.77\\
4&-115~nm  &-153~nm&-0.85&-0.88\\
\hline
\end{tabular}
\end{center}
\caption{Amplitude of poked actuators.}
\label{table1}
\end{table}

The first systematic component of the phase error 
indicates an ambiguity of the phase solution that should be explained.  
 This  kind  of ambiguity is observed when the focal-plane image can
be decomposed in two or more components that either do not interfere with each other or the
interference  terms are negligible. It happens, for example, in a practically
interesting case  where one of the components is produced by a multiplicative  pupil-plane phase grating. The diffractive orders formed by the grating interfere with
background speckles produced by   another   component. However,
in many cases, the brightness  of the background speckles is negligible in
comparison with the diffractive orders brightness, so the interference
term can be considered as equal to zero. As a result, all
phase solutions where the phase grating is arbitrarily shifted in the pupil plane
produce undistinguished focal-plane intensity distributions meaning that the phase
retrieval have an infinite number of ambiguous solutions. The interferometric
term, of course, is not equal to zero precisely.   However, any random registration noise 
makes solutions with arbitrary location of the grating in the pupil indistinguishable.
The phase ambiguity problem is additionally complicated by factors such as: 

(a) conjugated  phase solutions;

(b) small amplitude of DM diffractive peaks. The relative contrast of the DM diffraction peaks is less that $10^{-3}$. These peaks can reduce the Strehl ratio of the focal-plane PSF no more than by 0.02 which is comparable with Strehl ratio degradation due to the fiducial DM pattern. The  local convergence condition discussed in Section~\ref{methodc} is equivalent to the continuity condition for sequential pupil-plane phase probes.  Failure to converge with respect to the DM grating position could mean that phase probe changes that successfuly satisfy the continuity condition for the fiducial DM pattern case do not necessarily satisfy the continuity condition with respect to the DM surface grating. The number of degrees of freedom required to describe the DM grating is $\sim 200$ times larger than needed for a fiducial pattern that produces the same Strehl ratio degradation.  Thus, with respect to the DM grating the applied sequential pupil-plane phases are about $\sim 14$ times less ``continuous'' than with respect to the fiducial actuators pattern.  Consequently, a higher correlation rate between sequential random phase probes and a larger total number of iterations is required to recover phase features responsible for diffractive order formation. 

(c) Despite the high thermal and mechanical stability of the optical setup \cite{Belikov_2010}, slight long-term drift changes relative position of the optical elements. The drift produces beam-walk that results in temporal variations in pupil-plane phase affecting the phase retrieval algorithm. This drift thus causes uncertainty in the position of the observed irregular vertical phase structure (associated with OAP polishing features). 

 (d) Optical beams that form the PSF central lobe  and the DM diffractive orders are reflected from different parts of OAP2 which produces non-common path errors.Together with thermal and mechanical temporal instabilities these non-common path errors can affect the parameters of the recovered DM surface grid.

(e) Finally, an uncertainty in the size of the pupil  can be responsible for slightly different wavefront location  across the pupil in different retrieval runs.

All these factors contribute in the final phase difference map, but it is challenging to estimate their individual  contributions. Though the amplitude of the DM phase grating matches the amplitude measured interferometrically, the relative position of the DM grating remains uncertain. This uncertainty can probably be solved using the previously discussed defocus based diversity procedure. 

\section{Discussion}
\label{discussion}
 The proposed random phase probes approach can be used to test any optical system in which random correlated phase aberrations can be produced. 
 Such probes can be created, for example, by a DM, slow air turbulence or convection, temperature gradients in the system, or random relative motion of the different optical elements. Convergence of the algorithm solution to the real system wavefront is expected if the amplitude of the aberration creation process can be gradually decreased to 0. The method can be used for both large (as large as a few wavelength) and small optical aberrations. It is especially promising in the case when optical system performance is limited by non-common path propagation errors. The method can provide accurate, high resolution wavefront sensing not influenced by additional optical elements and does not require well-calibrated camera travel along the focus direction as with most phase diversity methods.

A list of possible method applications includes:

\noindent {\bf (1) Coronagraphy.} 
For a complex optical system with active wavefront control, the wavefront at different planes of the optical layout can be directly measured by using
the random phase probe approach. As a result the complete, high-accuracy system model can be produced. 
The direct imaging of exoplanets requires an extraordinary
high-contrast capability in the imaging system. 
Current starlight suppression systems based on a combination
of coronagraphy and wavefront control are able to
reach the $10^{-9}$--$10^{-10}$ broadband imaging contrast needed
for direct exoplanet studies. State of the art wavefront
control algorithms, such as EFC \cite{Giveon_2007, Krist_2015}, provide good local convergence.
However, they need a highly accurate model of
the starlight suppression system to reach the
necessary system performance. Random phase probes
based algorithms could provide such a model. These algorithms  
are fast, accurate and free from non-common
path errors that can affect the system performance. They
do not need to reconfigure the system for measurements  and
can be run remotely to perform system calibration
with an existing optical setup during space missions.

\noindent {\bf (2) Adaptive optics.} 
Obtained pupil-plane phase estimation
can be used for adaptive wavefront
correction, particularly to correct
wavefront aberrations caused by atmosphere turbulence.
Even in low flux conditions as low as 1-2 photons/speckle
the discussed approach allows
diffraction-limited wavefront correction with Strehl
ratios of 0.2-0.5. Taking into account the presented simulation results and 
assuming an atmosphere coherence time of 20 milliseconds (in visual spectral range) \cite{Ebersberger_1985,Fried_1987, Vernin_1991} the imaging 
frequency needed for the adaptive wavefront correction
can be roughly estimated  in the range of several hundred to one thousand Hz.

\begin{figure}[t]
\begin{center}
\begin{tabular}{c}
\includegraphics[width=0.9\hsize]{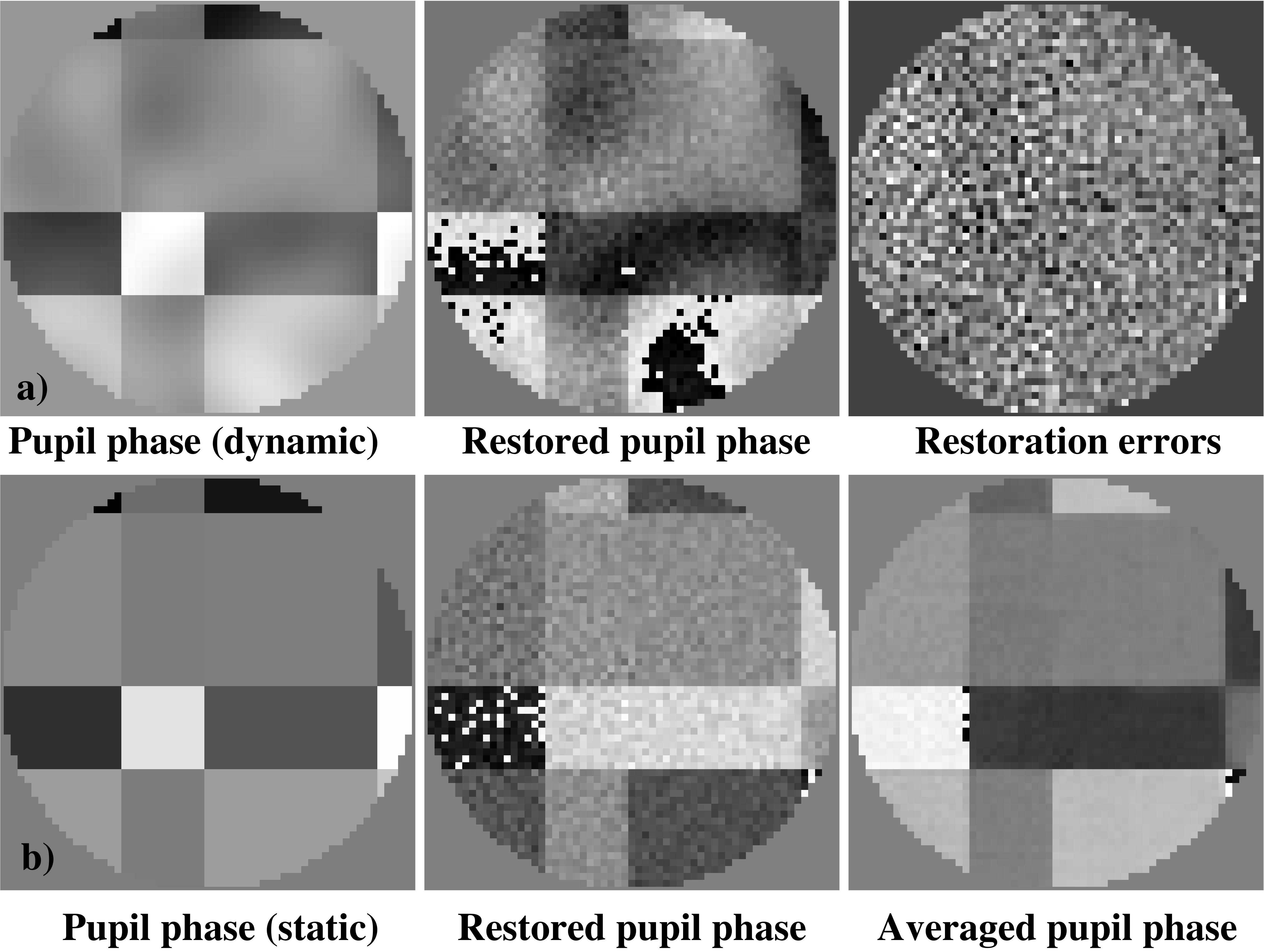}
\end{tabular}
\end{center}
\caption[fig20]
{\label{fig20}
The pupil-plane phase restoration in the case of segmented aperture. The reconstructed pupil-plane phase maps for the cases of dynamically changing (top)  and static (bottom) phase aberrations are shown. To visualize phase restoration errors for dynamically changing aberrations, a bias of 3$\sigma$ is added to the unwrapped error map such that the darkest and brightest pixels correspond to $\pm3\sigma$ levels.}
\end{figure}

\noindent {\bf (3) Segment co-phasing.} The random phase probe approach can be applied to both continuous and discontinuous pupil-plane phase distributions. Thus, the proposed wavefront reconstruction method can be used to co-phase sub-apertures of large segmented mirrors \cite{Nelson_2013}. A simulated demonstration of applicability of the wavefront reconstruction method to segment co-phasing is shown in Fig.~\ref{fig20}. The sample aperture consists of 15 different rectangular static segments with the piston rms of~ $\lambda/2$. In the presence of 1\% focal-plane background noise, the restored pupil-plane phase rms is about $\lambda/20$ corresponding to wavefront correction with Strehl ratio of about 0.92. For static aberrations the averaging procedure using 20 independent phase estimates can additionally improve the Strehl ratio to 0.993.  In the case where the segments are randomly moving relative to each other the procedure can provide the signal needed to dynamically correct  the segment positions. This position, of course, will be affected by integer-wave ambiguities visible in the restored wavefronts (Fig.~\ref{fig20}), that cannot be resolved in monochromatic Gerchberg-Saxton implementation of the considered approach. However, random phase probing  can be combined with any known broadband estimator (for example, with a gradient descent broadband estimator). In such combination the random phase probing would be responsible for the algorithm convergence while the broadband estimator would resolve the integer-wave ambiguities issue.
 Development of such estimators is beyond the scope of this paper.

\section{Conclusion}

Our method allows wavefront estimation in a pupil plane of an optical instrument by analyzing the effects of random (unknown) wavefront perturbations on a focal-plane image. Unlike other pupil-plane diversity methods, it has the advantage that it does not require knowledge of the applied pupil-plane perturbations, which makes it robust to model errors. In fact, the method can estimate the perturbations themselves.

We have considered the possibility of the appearance of ambiguous solutions and have proven the uniqueness theorem. We have also discussed the problem of global convergence that can be solved the defocus based phase divergence that provided the global solutions in all our simulations. 

For phase retrieval, we have demonstrated low sensitivity to photon noise and registration noise. To increase  accuracy of the final phase estimates an averaging procedure has been proposed.  The considered algorithms provide fast and reliable pupil-plane phase measurements for both large (on the order of a few $\lambda$) and small (optical surface structures as shallow as 30-40 nm can be measured) phase aberrations. 
The algorithm can be applied to both continuous and discontinuous pupil-plane phase distributions and can be used to measure either static and or dynamically changing pupil-plane phases. Similar to other focal-plane wavefront sensing methods, the random phase probe estimate method uses common-path and can thus be used by applications whose performance is limited by non-common path errors. 

The phase retrieval solutions can be affected by systematic errors relevant especially to periodical high frequency pupil-plane structures whose position cannot be efficiently detected by applied phase probes. In this case an appropriate solution can be obtained by increasing defocus term in the used phase diversity procedure that increases the related interferometric term.



For implementation of the random phase probe method, the following requirements should be met:
\begin{description}
\item [--] the focal-plane imaging procedure should provide an adequate image sampling;
\item [--] a capability to create  random  pupil-plane phase aberrations and control
      their amplitudes using (for example) a DM. Note, that the particular
      location of the DM is not important. The DM can be placed anywhere
      in the optical system except at the system focus.
\item [--] the imaging frequency should be high enough to maintain
      high sequential correlation of the focal-plane images in the case when
      the random phase probes are created by a rapid dynamical process such as
      the atmosphere turbulence or convection.
\item [--] an appropriate propagation model should be used, that assumes that the
procedure can be used in the case of Fresnel diffraction for example.
\end{description}

The described procedure can be implemented for a wide class of optical applications, including those that use the DM for adaptive wavefront correction.

Finally, the system noise not only limits the system performance, sometimes it helps to improve the performance.

\vskip 1em

{\it\noindent This work was funded by the NASA Ames FY17 Center Innovation Fund (CIF) program. We thank the two referees and the editor for constructive comments that significantly improved the manuscript.}

\bibliography{report}   

\end{document}